\documentclass[twocolumn]{aastex631}
\usepackage{bm}
\usepackage{colortbl}
\usepackage{appendix}
\usepackage{float}

\def\eg{e.g.,}

\newcommand{\redchi}{${\chi}_{\nu}^2$}
\newcommand{\sqdeg}{deg$^{2}$}
\newcommand{\sqam}{arcmin$^{2}$}

\newcommand{\numlaecands}{920,715} 
\newcommand{\numoIIcands}{451,224} 
\newcommand{\numoverlap}{28,705} 
\newcommand{\numredshifts}{9,710} 
\newcommand{\numdiagnose}{6,480} 
\newcommand{\numdiagnosearchive}{1,701} 
\newcommand{\numunknowndiagnose}{21,081} 
\newcommand{\numbadspectra}{998} 
\newcommand{\numcatmatch}{7,409} 
\newcommand{\numnewredshifts}{9,087} 
\newcommand{\OII}{[\ion{O}{2}]} 
\newcommand{\OIII}{[\ion{O}{3}]}
\newcommand{\NeIII}{[\ion{Ne}{3}]}
\newcommand{\numshimwell}{325,694} 
\newcommand{\numradio}{318,520} 

\newcommand{\numphotz}{11,807} 
\newcommand{\numspecz}{2,690} 
\newcommand{\numcounter}{231,716} 

\newcommand{\nagn}{48,194} 
\newcommand{\nstar}{775,063} 
\newcommand{\nlzg}{98,801} 
\newcommand{\nstars}{197} 
\newcommand{\nagns}{804} 
\newcommand{\nlowzgal}{6,394} 
\newcommand{\nhighzgal}{1,075} 
\newcommand{\narchive}{757} 
\newcommand{\numfiberspec}{523 million} 
\newcommand{\hetdexspringarea}{67.48} 
\newcommand{\sfrL}{SFR-L$_{\rm{150MHz}}$}
\usepackage{gensymb}

\begin{document}

\title{HETDEX-LOFAR Spectroscopic Redshift Catalog 
\footnote{Based on observations obtained with the Hobby-Eberly Telescope, which is
a joint project of the University of Texas at Austin, the Pennsylvania State
University, Ludwig-Maximilians-Universit\"at M\"unchen, and
Georg-August-Universit\"at G\"ottingen.}
}

\author[0000-0002-1998-5677]{Maya H. Debski}
\affiliation{Department of Astronomy, The University of Texas at Austin, 2515 Speedway Boulevard, Austin, TX 78712, USA}

\author[0000-0003-2307-0629]{Gregory R. Zeimann}
\affiliation{Hobby Eberly Telescope, University of Texas at Austin, 2515 Speedway Boulevard, Austin, TX, 78712, USA}
\affiliation{McDonald Observatory, The University of Texas at Austin, 2515 Speedway Boulevard, Austin, TX 78712, USA}

\author[0000-0001-6717-7685]{Gary J. Hill} 
\affiliation{Department of Astronomy, The University of Texas at Austin, 2515 Speedway Boulevard, Austin, TX 78712, USA} 
\affiliation{McDonald Observatory, The University of Texas at Austin, 2515 Speedway Boulevard, Austin, TX 78712, USA}

\author[0000-0001-7240-7449]{Donald P. Schneider}
\affiliation{Department of Astronomy \& Astrophysics, The Pennsylvania
State University, University Park, PA 16802, USA}
\affiliation{Institute for Gravitation and the Cosmos, The Pennsylvania State University, University Park, PA 16802, USA}

\author[0000-0003-0487-6651]{Leah Morabito}
\affiliation{Centre for Extragalactic Astronomy, Department of Physics, Durham University, South Road, Durham DH1 3LE, UK}
\affiliation{Institute for Computational Cosmology, Department of Physics, Durham University, South Road, Durham DH1 3LE, UK}

\author[0000-0002-3031-2588]{Gavin Dalton}
\affiliation{Oxford Astrophysics, University of Oxford, Keble Road, Oxford
OX1 3RH, UK}
\affiliation{RALSpace, STFC Rutherford Appleton Laboratory, Didcot, OX11 0QX, UK}

\author[0000-0001-7039-9078]{Matt J.~Jarvis}
\affiliation{Oxford Astrophysics, University of Oxford, Keble Road, Oxford
OX1 3RH, UK}

\author[0000-0002-2307-0146]{Erin Mentuch Cooper}
\affiliation{Department of Astronomy, The University of Texas at Austin, 2515 Speedway Boulevard, Austin, TX 78712, USA}
\affiliation{McDonald Observatory, The University of Texas at Austin, 2515 Speedway Boulevard, Austin, TX 78712, USA}

\author[0000-0002-1328-0211]{Robin Ciardullo} \affiliation{Department of Astronomy \& Astrophysics, The Pennsylvania State University, University Park, PA 16802, USA} \affiliation{Institute for Gravitation and the Cosmos, The Pennsylvania State University, University Park, PA 16802, USA}

\author[0000-0003-1530-8713]{Eric Gawiser}
\affiliation{Department of Physics and Astronomy, Rutgers, the State University of New Jersey, Piscataway, NJ 08854, USA}

\author[0000-0003-4270-5968]{Nika Jurlin}
\affiliation{Department of Astronomy, The University of Texas at Austin, 2515 Speedway Boulevard, Austin, TX 78712, USA}

\begin{abstract}
We combine the power of blind integral field 
spectroscopy from the Hobby-Eberly Telescope (HET) Dark Energy Experiment (HETDEX) with sources detected by the Low Frequency Array (LOFAR) to construct the HETDEX-LOFAR Spectroscopic Redshift Catalog. Starting from the first data release of the LOFAR Two-metre Sky Survey (LoTSS), including a value-added catalog with photometric redshifts, we extracted \numoverlap\ HETDEX spectra. Using an automatic classifying algorithm, we assigned each object a star, galaxy, or quasar label along with a velocity/redshift, with supplemental classifications coming from the continuum and emission line catalogs of the internal, fourth data release from HETDEX (HDR4). We measured \numnewredshifts\ new redshifts; in combination with the value-added catalog, our final spectroscopic redshift sample is \numredshifts\ sources. This new catalog contains the highest substantial fraction of LOFAR galaxies with spectroscopic redshift information; it improves archival spectroscopic redshifts, and facilitates research to determine the \OII\ emission properties of radio galaxies from $0.0 < z < 0.5$, and the Ly$\alpha$ emission characteristics of both radio galaxies and quasars from $1.9 < z < 3.5$. Additionally, by combining the unique properties of LOFAR and HETDEX, we are able to measure star formation rates (SFR) and stellar masses. Using the Visible Integral-field Replicable Unit Spectrograph (VIRUS), we measure the emission lines of [\ion{O}{3}], [\ion{Ne}{3}], and \OII\ and evaluate line-ratio diagnostics to determine whether the emission from these galaxies is dominated by AGN or star formation and fit a new \sfrL\ relationship.
\end{abstract}

\section{Introduction} \label{sec:intro}
For several decades, extragalactic radio surveys were a powerful probe of the distant Universe. In fact, until the mid-1990s, they served as an effective method for finding high redshift galaxies through optical identification of ultra steep radio sources, a poorly understood type of diffuse radio source characterized by power law spectra \cite[\eg][]{Slee01,Feretti12,Whyley24}. Recent radio surveys have reached sub-milliJansky flux density levels, providing the framework to make these surveys a means of identifying star-forming galaxies. Previous studies have demonstrated a tight correlation between low-frequency radio continuum, which is dominated by synchrotron emission of relativistic electrons produced by supernovae, and the far-infrared (FIR) flux of galaxies \cite[\eg][]{Yun01}. FIR emission is an established strong indicator of star formation rate (SFR) \citep{Yun01}, thereby legitimizing the idea that radio continuum can act as a tracer of SFR \cite[\eg][]{Condon02,panella09,Heesen14,Davies2017,Smith21}. 

Radio surveys have played a small part in investigating star formation history, as they are generally limited by their sensitivity; however, star forming galaxies become increasingly important at fainter flux densities and totally dominate the source counts below $\simeq$ 0.1 mJy at 1.4 GHz \cite[\eg][]{padovani11,padovani15,bonzini13}, opening the door for new, more sensitive radio surveys to push to the forefront of SFR investigations (\citealp{DeZotti19}). In particular, the introduction of the Square Kilometer Array (SKA; \citealp{Grainge17}) with its square kilometer collecting surface and large range of frequencies (between 50 MHz - 24 GHz) will extend the flux density limit more than three orders of magnitude. The science of the SKA will be used to explore areas such as strong-field tests with pulsars and black holes, cosmic dawn and the epoch of reionization, cosmology and dark energy, the origin and evolution of cosmic magnetism, galaxy evolution probed by neutral hydrogen, the cradle of life and astrobiology, and galaxy and cluster evolution (\citealp{DeZotti19}). To prepare for this new era of radio surveys, SKA will work alongside Pathfinders, like the International Low-frequency Array (LOFAR) Telescope (ILT), to contribute scientific and technical developments for direct use by SKA. 

The combination of multiple massive surveys at different wavelengths enables scientific projects otherwise unachievable \cite[\eg][]{Smith16,Williams19}. The LOFAR Two-metre Sky Survey (LoTSS) is a sensitive, high-resolution (120-168 MHz, centered at 150 MHz) survey that has already collected millions of sources and is advancing our understanding of the formation and growth of massive black holes \cite[\eg][]{Best14,Mingo19,Mingo22,Yue23,Sabater19}, the evolution of galaxy clusters \cite[\eg][]{Venemans07,Wylezalek13,Timmerman22,Timmerman24,Botteon20}, and the properties of high redshift radio sources \cite[\eg][]{Gloudemans21,Cordun23}. However, many of these scientific forefronts require optical counterparts for multi-wavelength matching as well as a robust set of distances or redshifts.
 
HETDEX \citep{gebhardt2021, hill2021} is blind-spectroscopic survey conducted with the wide-field upgraded Hobby Eberly Telescope \citep{Ramsey1998, hill2021} using the Visible Integral Field Replicable Unit Spectrograph (VIRUS; \citealp{hill2021}).\footnote{VIRUS is a joint project of the University of Texas at Austin, Leibniz-Institut f{\" u}r Astrophysik Potsdam (AIP), Texas A\&M University (TAMU), Max-Planck-Institut f{\" u}r Extraterrestriche-Physik (MPE), Ludwig-Maximilians-Universit{\" a}t M{\" u}nchen, Pennsylvania State University, Institut f{\" u}r Astrophysik G{\" o}ttingen, University of Oxford, Max-Planck-Institut f{\" u}r Astrophysik (MPA), and The University of Tokyo.} HETDEX aims to measure the expansion history of the Universe at $z\simeq 2.5$ by detecting and mapping the spatial distribution of about a million Ly$\alpha$ emitting galaxies (LAEs). The redshift range for LAE detection is $1.9 < z < 3.5$ over a total of $\sim540$ \sqdeg\ (11 Gpc$^3$ comoving volume) including $\sim400$ \sqdeg\ in the HETDEX Spring field. The internal data release 4 includes \hetdexspringarea\ \sqdeg\ of spectroscopic observations in the Spring Field with wavelengths spanning 3470-5540 \AA. This amounts to \numfiberspec\ fiber spectra. In the LOFAR/HETDEX overlap region, we aim to produce a new, large spectroscopic radio-source catalog that would facilitate breakthroughs in the study of galaxy proto-clusters, emission line properties of radio galaxies, and even radio-loud stars.

The combination of wide-field photometric surveys for spectral energy distribution information and optical spectroscopic follow-up from HETDEX/VIRUS offers a characterization of key physical parameters, the most important of which being distance as represented by spectroscopic redshift \cite[\eg\ SDSS;][]{almeida23}. The combination of HETDEX and LOFAR will also allow the measurement of SFR, and stellar mass because of the VIRUS sensitivity. Both the LOFAR radio selection and HETDEX optical spectroscopic identifications are sensitive to SFR and AGN activity. Using VIRUS, we can measure the emission lines of \OIII, H$\beta$, \NeIII, and \OII\null. Emission line diagnostics like those of \OIII/H$\beta$, \NeIII/[O~II], R$_{23}$, and \OIII/\OII\ can provide information on a system's metallicity and ionization parameter, and discriminate between excitation by AGN or star formation.  Moreover, by comparing these line ratios as a function of stellar mass with those determined for galaxies found via selection methods, we can ultimately investigate how similar or dissimilar populations of galaxies are.
 
In $\S$\ref{sec:data}, we describe the observations, data sets, and tools required to identify HETDEX counterparts to LOFAR sources and analyze the objects $\S$\ref{sec:zanalysis} focuses on the redshift analysis of the sample, including a description of the classification code Diagnose that was developed for the HET VIRUS Parallel Survey (HETVIPS) and its application in this catalog. We also describe the matching process for the sources within both the HETDEX survey and LoTSS, as well as the criteria for determining whether the matches were accurate. This section also features a broad overview of the redshift results alongside the overall catalog breakdown and selected results from the catalog. In addition, we examine star formation in a sub-sample of galaxies in $\S$\ref{sec:discussion}. Finally, we discuss possible scientific applications and uses for this data set in $\S$\ref{sec:summary}. Throughout this work, we use flat $\Lambda$CDM cosmological parameters $H_0 = 67.66$ km Mpc$^{-1}$ s$^{-1}$ and $\Omega_{m,0} = 0.30966$ \citep{planck18}.

\section{Data and Observations}
\label{sec:data}
In this section, we present an overview of the LoTSS DR1 and the HDR4 data sets used in this work. This includes the value-added catalogs from previous efforts, the matching methodology, and the spectral extractions.

\subsection{LOFAR Two-metre Sky Survey}
The LOFAR Two-metre Sky Survey (LoTSS) is a high-resolution 120-168 MHz survey centered at 150 MHz, with a median sensitivity of $S_{\text{150 MHz}} = 71 \mu \text{Jy}\ \rm beam^{-1}$ and a point-source completeness of 90\% at an integrated flux density of 0.45 mJy \citep{Shimwell17}. The spatial resolution of the images is 6\arcsec\ and the astrometric accuracy of the data is within $0.2$\arcsec\ \citep{Shimwell19}. The first data release includes 424 \sqdeg\ in the HETDEX Spring Field (RAs between 10$^{\text{h}}$45$^{\text{m}}$ and 15$^{\text{h}}$30$^{\text{m}}$ and DECs ranging from 45\degree00\arcmin to 57\degree00\arcmin). There are \numshimwell\ sources in the first data release for the HETDEX Spring field. 
Additionally, the LoTSS DR1 provides the astrometric precision 
needed to identify optical and infrared counterparts. We use the first data release because it specifically targeted the HETDEX Spring field, and while LoTSS DR2 (formed by two regions centered at RA = 12h45m00s, DEC = +44$^\circ$30$'$00$''$ and RA = 1h00m00s, DEC = +28$^\circ$00$'$00$''$, spanning 4178 and 1457 \sqdeg\ respectively) is available, it only expands upon the area of the first data release, essentially making DR1 and DR2 interchangeable for our purposes. 

\cite{Williams19} combined Pan-STARRS $grizy_{\rm P1}$ photometry \citep{cham2016} with data from the Wide-field Infrared Survey Explorer (WISE; \citealp{Wright2010}) over the LoTSS DR1 region. Using a combination of statistical techniques and visual identification, \cite{Williams19} constructed a color- and magnitude-dependent likelihood ratio method for statistical identification. This resulted in a value-added catalog\footnote{\url{https://lofar-surveys.org/public/LOFAR\_HBA\_T1\_DR1\_merge\_ID\_optical\_f\_v1.2b\_restframe.fits}} with \numradio\ radio sources, of which \numcounter\ (73\%) have optical and/or IR identifications in Pan-STARRS and WISE.

Along with optical/IR counterparts for each radio source, the LoTSS DR1 value-added catalog includes photometric redshift estimates from \cite{Duncan19}. These estimates are crucial for identifying properties of the radio sources, as SDSS provides spectroscopic redshifts for less than one percent (\numspecz) of the sources. In the near future, the William Herschel Telescope Enhanced Area Velocity Explorer (WEAVE; \citealp{Dalton12,Dalton14}) multi-object and integral field spectrograph will measure redshifts of over a million LoTSS sources as part of the WEAVE-LOFAR survey \citep{Smith16}. We can take the first step in informing that large effort by increasing the known spectroscopic redshifts by combining the LoTSS DR1 value-added catalog with the fourth data release from the HETDEX survey.

\subsection{HETDEX Data Release 4}
The HETDEX survey is designed to measure the Hubble expansion parameter and angular diameter distances by using the spatial distribution of nearly one million Ly$\alpha$ emitting galaxies. The survey employs the VIRUS instrument which is comprised of a set of 78 fiber integral field units (IFUs) feeding 156 identical spectrographs that produce 34,944 spectra covering the wavelength range 3470\,\AA\ -- 5540\,\AA\ at a resolving power $R \approx 800$. The IFUs are arrayed in a grid pattern on the sky with a fill factor that is $\simeq$ 1/4.5, covering 56 \sqam\ within an 18\arcmin\ diameter field. Each IFU covers a solid angle of approximately $51\arcsec \times 51\arcsec$ and feeds two spectrographs, each with 224 fibers. The individual fibers are $1\farcs 5$ in diameter and the spacing between the fiber centers is $2\farcs 2$. During HETDEX observations, a dither pattern of three exposures nearly fills these gaps ($\sim$94\% sky coverage;
see \cite{hill2021} for details). The astrometric accuracy of the fiber positions for HETDEX is 0.35\arcsec \citep{gebhardt2021}.

The HETDEX survey serves as the primary observing mode at the HET during dark sky conditions, accumulating a wealth of data. The internal fourth data release completed observations on 2023-08-31, and it includes $\sim$67.48 \sqdeg\ of fiber sky coverage with exposures from August 2017 through August 2023. The majority of these observations are in the HETDEX Spring Field with \hetdexspringarea\ \sqdeg\ sky coverage and \numfiberspec\ fiber spectra. The data reductions are described in \citet{gebhardt2021}, but to summarize:  the HETDEX team produced two main products --- a full set of flux-calibrated fiber spectra and a catalog of automatically detected and classified sources \citep{Cooper23}.

To construct the source catalog the HETDEX team ran two object detection algorithms: one designed to find emission line sources and the other built to search for continuum emission \citep{Cooper23}. From these two raw catalogs, source sizes were defined using a friends of friends algorithm to avoid multiple detections of the same object. The team then took a multi-pronged approach to source classification and redshift assignment. The details of the classification and redshift assignment can be found in \citet{Cooper23}. In short, each source was classified either as a star (STAR), a low redshift galaxy with no \OII\ emission (LZG), an \OII\ emitting galaxy (O II), a Ly$\alpha$ emitting galaxy (LAE), or an active galactic nuclei (AGN). 

The HDR4 catalog is dominated by emission-line galaxies and includes \numlaecands\ LAE candidates with $1.88<z<3.52$ and with a signal-to-noise greater than 4.8. Also included in the catalog are \numoIIcands\ \OII-emitting galaxies at $z<0.5$, \nstar\ stars, \nlzg\ low-redshift ($z<0.5$) galaxies without emission lines, and \nagn\ AGN\null. The catalog provides sky coordinates, redshifts, line identifications, classification information, line fluxes, \OII~and Ly$\alpha$ line luminosities when applicable, and spectra for all identified sources processed by the HETDEX detection pipeline. 

Although the catalog provides many of the products that we need, we can supplement the HETDEX effort by extracting spectra at the precise locations of the LoTSS sources, and then running similar classification tools for a complete redshift analysis of the combined catalogs.
 
\subsubsection{HETDEX Spectral Extractions}

We use the HETDEX-API\footnote{\url{https://github.com/HETDEX/hetdex\_api}} (\citealp{Cooper23}) to extract a spectrum for the \numoverlap\ sources in the LoTSS DR1 catalog with fiber coverage in HDR4. For the \numoverlap\ sources, we first collect all fiber spectra within a $3.5\arcsec$ radius. Using the seeing measured from the VIRUS data (see \citealp{gebhardt2021} for details), we construct a Moffat PSF model ($\beta = 3.5$, \citealt{Moffat1969}). At each wavelength, we shift our fiber locations following the differential atmospheric refraction models for the fixed-altitude HET and convolve the PSF with the VIRUS fibers. This calculates the fraction of the object's light covered by each fiber. Using these fiber coverage values as weights, we normalize the weights to one, retaining the normalization value, and perform a weighted extraction using the \citet{Horne1986} optimal extraction formula. Finally, the resultant spectrum is corrected to a total flux using the normalization value. Within a $3\farcs 5$ aperture, the total fiber coverage is between 90-95\%.

Although a PSF spectral extraction is not a natural method for each source in the catalog, it provides a higher signal to noise methodology than simpler aperture extractions and only introduces a minimal chromatic flux response issue for extended sources. We are not immediately concerned with the absolute calibration of our spectral extractions as the primary goal is the determination of redshift. A more appropriate extraction can be done for individual science cases starting from the information driven by the description below.

The vast majority of our sample have an average continuum signal to noise (S/N) of less than two per 2\,\AA\ pixel in the wavelength window of 4670-4870\AA\null. However, it is clear that if the S/N is greater than $\sim 4$, a clear source classification and redshift measurement are possible by eye. Although the sample is not too large for visual analysis, automatic tools with repeatable and similar success rates are available for this purpose \citep[\eg][]{bolton2012}.

\section{Classifications and Redshifts}
\label{sec:zanalysis}

The methodology for our classification scheme begins with the spectral extractions at the sky positions of each LoTSS DR1 source. We use Diagnose \citep{Debski24}, a spectral classification code developed for the HET VIRUS Parallel Survey (HETVIPS) catalog \citep{Zeimann24}, to automatically classify each source and assign a redshift.

\subsection{Diagnose}
The Diagnose code assigns one of four classifications for each source (star, galaxy, quasar, or unknown) while returning a redshift estimate for the galaxies and quasars and a velocity estimate for the stars. Diagnose determines a spectral classification and redshift estimate for each source via a ${\chi}^2$ minimization for linear combinations of principal component templates. In particular, Diagnose uses a principal component analysis (PCA) with the templates of redrock\footnote{\url{https://github.com/desihub/redrock-templates}}, which include 10 components for galaxies and four components for quasars. Stars are classified by type, with six components for B, A, F, G, K, M, and white dwarfs. 

By convolving the high resolution templates of redrock to the $R\approx 800$ resolution of VIRUS, and then fitting these templates to the data using a range of redshifts, Diagnose computes three best-fit ${\chi}^2$ values:  one for stellar type and velocity, one for galaxy type and redshift, and one for a quasar and redshift.  Diagnose then compares the best fit of these three \redchi\ values to the second best fit and evaluates the difference against a statistical threshold. If this difference is larger than a statistical threshold, the source is classified as the best fit template (i.e., star, galaxy, or quasar). If the difference is not larger than the threshold, the source is classified as unknown. Using Diagnose, the HETDEX-LOFAR Spectroscopic Redshift Catalog is able to produce classifications and redshift estimates for sources within LoTSS with no known spectroscopic redshifts.

The power of Diagnose is in identifying strong features, usually in the continuum.  As the sources become fainter with lower signal to noise in the continuum, Diagnose becomes less effective and often returns an a unknown label. However, many of the radio sources in the HETDEX-LOFAR catalog have strong emission features but weak continua.  For these sources, we can use the HDR4 catalog to both check our initial Diagnose classifications and supplement our identifications.

\subsection{LoTSS-HDR4 Catalog Matching}

\begin{figure}[t]
  \centering
  \includegraphics[width=1\columnwidth]{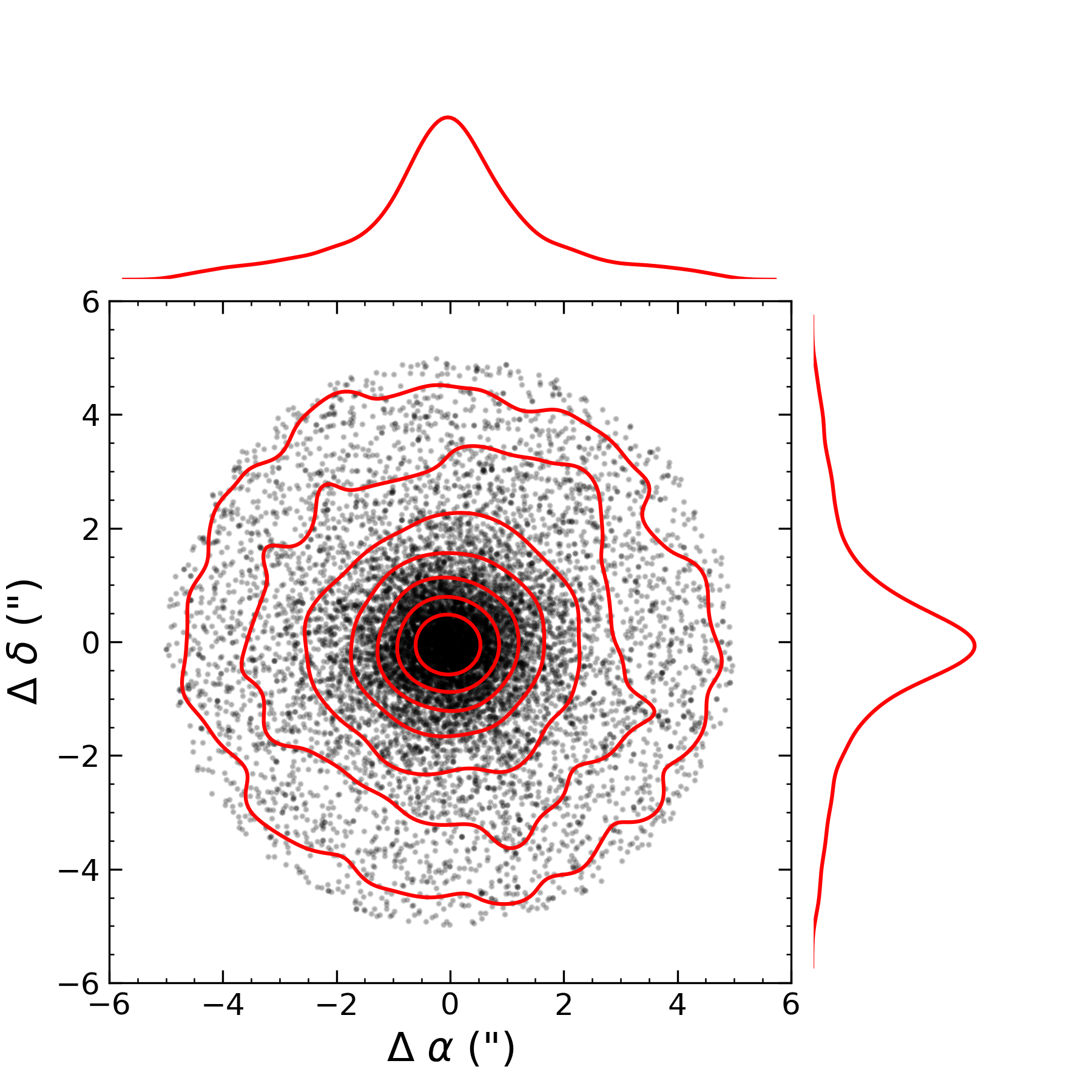}
  \caption{Comparison between $\Delta\alpha$ and $\Delta\delta$ at initial matching radius of 5\arcsec for the \numcatmatch\ matches found between the HETDEX HDR4 catalog and the LoTSS sources. The x-axis represent the $\Delta\alpha$ and the y-axis represents the $\Delta\delta$. Sources within a radius of 2\arcsec have less than 5.43\% chance of being false matches when fitted with a uniform + Gaussian distribution model.}
  \label{fig:radec}
\end{figure}

To match the LoTSS data set to sources in the HDR4 catalog, we use the HETDEX-API, including the ElixerWidget \citep{Davis23}. We began by using an initial matching radius of 5\arcsec\ to find counterparts for the \numoverlap\ LoTSS sources. Given the astrometric uncertainty of the two catalogs, this should be a more than ample starting threshold. We found \numcatmatch\ matches and plot the $\Delta$RA vs. $\Delta$Dec in Fig~\ref{fig:radec}. We fit a 2D uniform + Gaussian distribution model in $\Delta$RA and $\Delta$Dec space to determine the matching standard deviation and estimate the spurious match fraction. We found a standard deviation of $0.61$\arcsec\ in $\Delta$RA and $0.65$\arcsec\ in $\Delta$Dec. Thus, we use a final matching radius of 2\arcsec\, which is roughly 3 times the standard deviation. For a matching radius of 2\arcsec\ we estimate that only $\sim$5\% of the counterparts are spurious by subtracting the area under the Gaussian distribution model from the area under the 2D uniform distribution model and dividing by the area under the 2D uniform distribution model.

\subsection{Combining Diagnose and HDR4}
\label{sec:combine}

Here we describe how we combine our Diagnose classifications/redshifts with the classifications/redshifts obtained from the LoTSS-HETDEX catalog matching. Starting from the \numoverlap\ radio sources, Diagnose confidently identified \numdiagnose\ objects as a star, galaxy, or quasar; The remaining \numunknowndiagnose\ sources did not have a reliable classification. Of those, \numbadspectra\ objects had insufficient spectral coverage for either a Diagnose or a catalog label; this is due to masking of bad fibers/amplifiers in the dataset after the initial spatial matching. 

For the \numcatmatch\ spatial matches between LoTSS-HETDEX catalog, we collected the classification and redshift from HDR4. When comparing the sources with both a Diagnose and HDR4 redshift, we find good agreement, with 92.3\% of the objects agreeing to within $\Delta$z = 0.05. This is not entirely surprising as the HDR4 classification scheme uses Diagnose for sources with continuum $g$-band magnitudes brighter than 22.  For the objects with discrepant redshifts, the most common reason was Diagnose labeling an emission line as \OII, rather than Ly$\alpha$. This occurred 3.03\% of the time. As noted before, we also expect a $\sim$5\% spurious match fraction between HDR4 and LoTSS, which may account for the remaining disagreement between the two redshift estimates.

\begin{figure}[t]
  \centering
  \includegraphics[width=1\columnwidth]{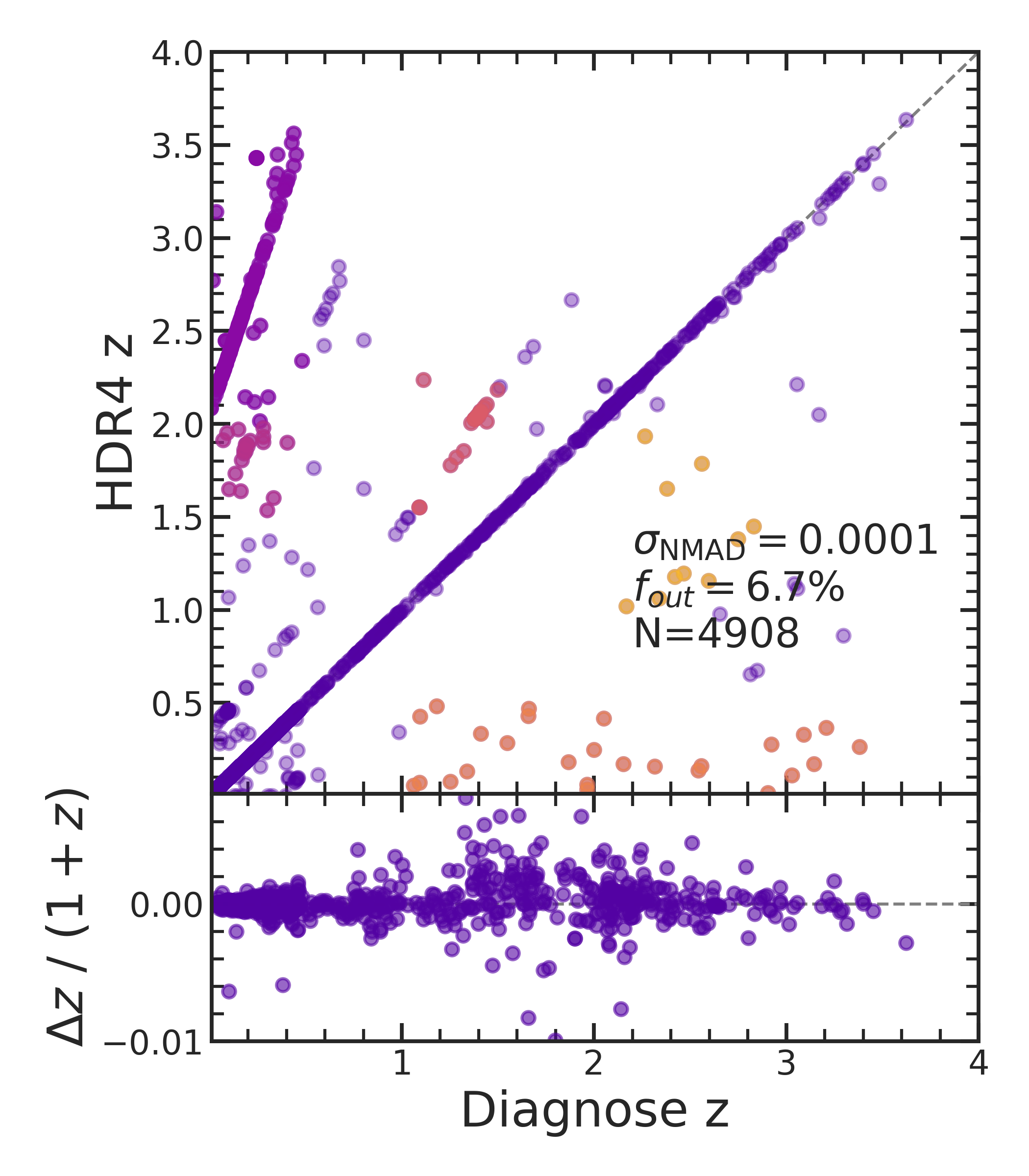}
  \caption{Comparison between sources with both HDR4 redshifts and Diagnose redshifts. There is good agreement between the two redshifts, though there is some scatter. The different colored groups represent areas of interest that were further investigated (see \S\ref{sec:combine}) to make the correct redshift assignments. The normalized median absolute deviation (NMAD) and outlier fraction are calculated as in \cite{momcheva16}.}
  \label{fig:hdr4_diagnose}
\end{figure}

Comparing the HDR4 spectroscopic redshifts with our Diagnose redshifts, we find 4,908 sources in common. Figure~\ref{fig:hdr4_diagnose} compares the two redshifts; the outlier fraction is 6.7\%, while the standard deviation of the remaining objects is \textrm{${\sigma}_z = 0.0001$}. Figure~\ref{fig:hdr4_diagnose} also highlights groups of particular interest within the set of objects with discrepant redshifts.  For each of these groups, we investigated the spectra by eye and determined criteria for determining which redshift is the best fit. These criteria and the process for choosing the best redshift are detailed in Appendix \ref{appendix:a}. After applying all of the criteria to the different groups of spurious matches, the outlier fraction reduces from 6.7\% to 2.3\%. 

\subsection{LoTSS DR1 Value-Added Catalog Redshifts}

Within \numoverlap\ LoTSS sources, there are \numphotz\ objects with photometric redshifts and \numspecz\ sources with spectroscopic redshifts in the value-added catalog from \citet{Williams19} and \citet{Duncan19}. The majority of spectroscopic redshifts were compiled from the Sloan Digital Sky Survey (SDSS) Data Release 14 \cite[DR14;][]{sdss14}. These redshifts were supplemented by additional spectroscopic data from a range of deep optical surveys in the literature, mostly covering the Extended Groth Strip within the HETDEX Spring field. We refer to the spectroscopic redshifts in the value-added catalog as archival redshifts.

\begin{figure}[t]
  \centering
  \includegraphics[width=\columnwidth]{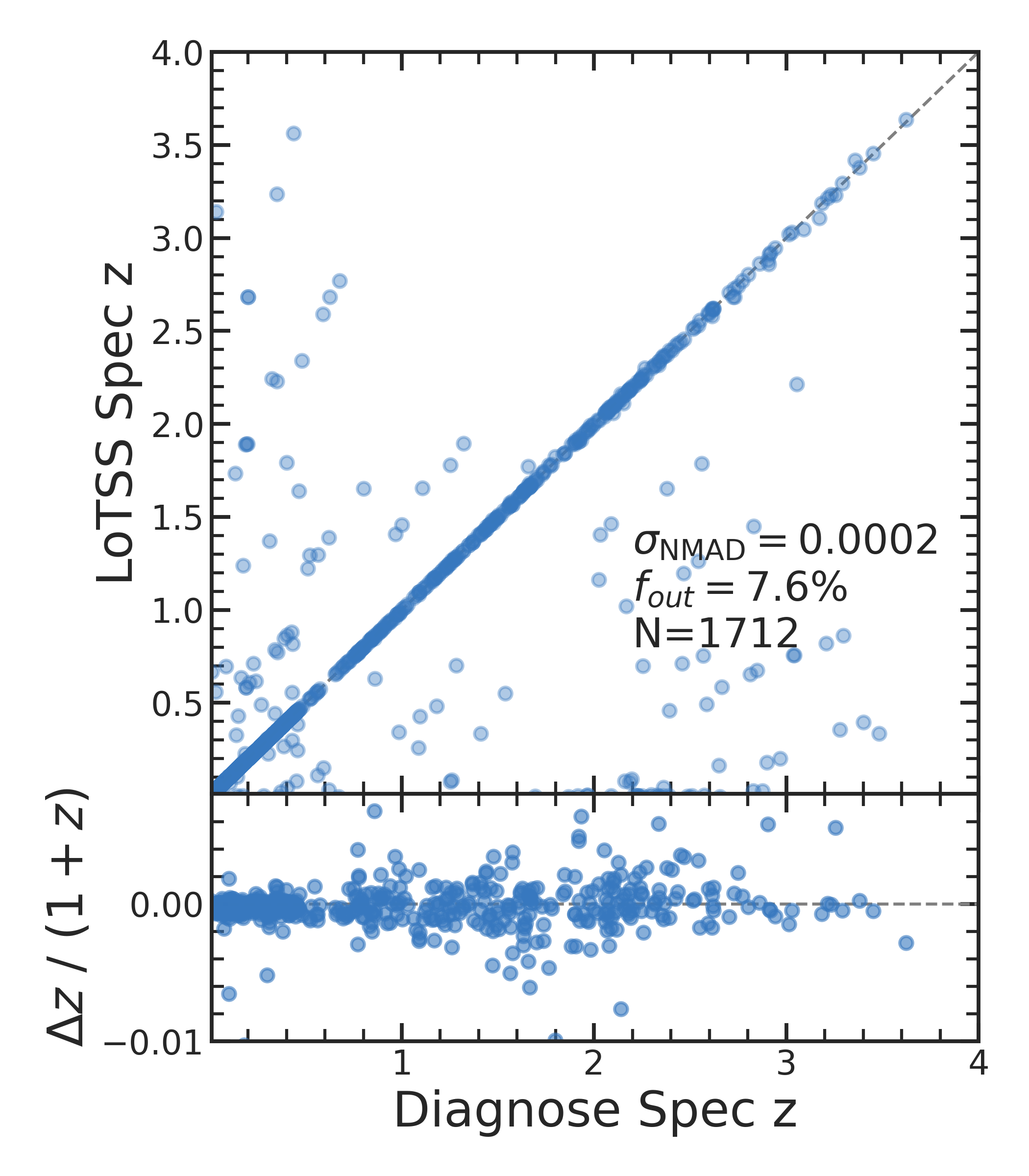}
  \caption{Comparison between LoTSS previously determined spectroscopic redshifts and Diagnose spectroscopic redshifts for the same sources. In total, there were 1,098 LoTSS sources with previous spectroscopic redshift counterparts. The normalized median absolute deviation (NMAD) and outlier fraction are calculated as in \cite{momcheva16}. The LoTSS redshifts and Diagnose redshifts are in good agreement. }
  \label{fig:knownspecz}
\end{figure}

Comparing the archival spectroscopic redshifts with our HETDEX-LOFAR catalog we find \numdiagnosearchive\ sources in common. The vast majority ($\sim$75\%) of the archival redshifts that are not in our catalog are in the redshift desert of VIRUS (0.5 $<$ \textit{z} $<$ 1.9) where there are no strong emission lines. Investigating the overlapping sources, we find good agreement between our spectroscopic redshifts and those in the literature. Figure~\ref{fig:knownspecz} shows the outlier fraction is 7.6\% with a standard deviation of non-outlying sources is \textrm{${\sigma}_z = 0.0002$}.

\begin{figure}[t]
  \centering
  \includegraphics[width=\columnwidth]{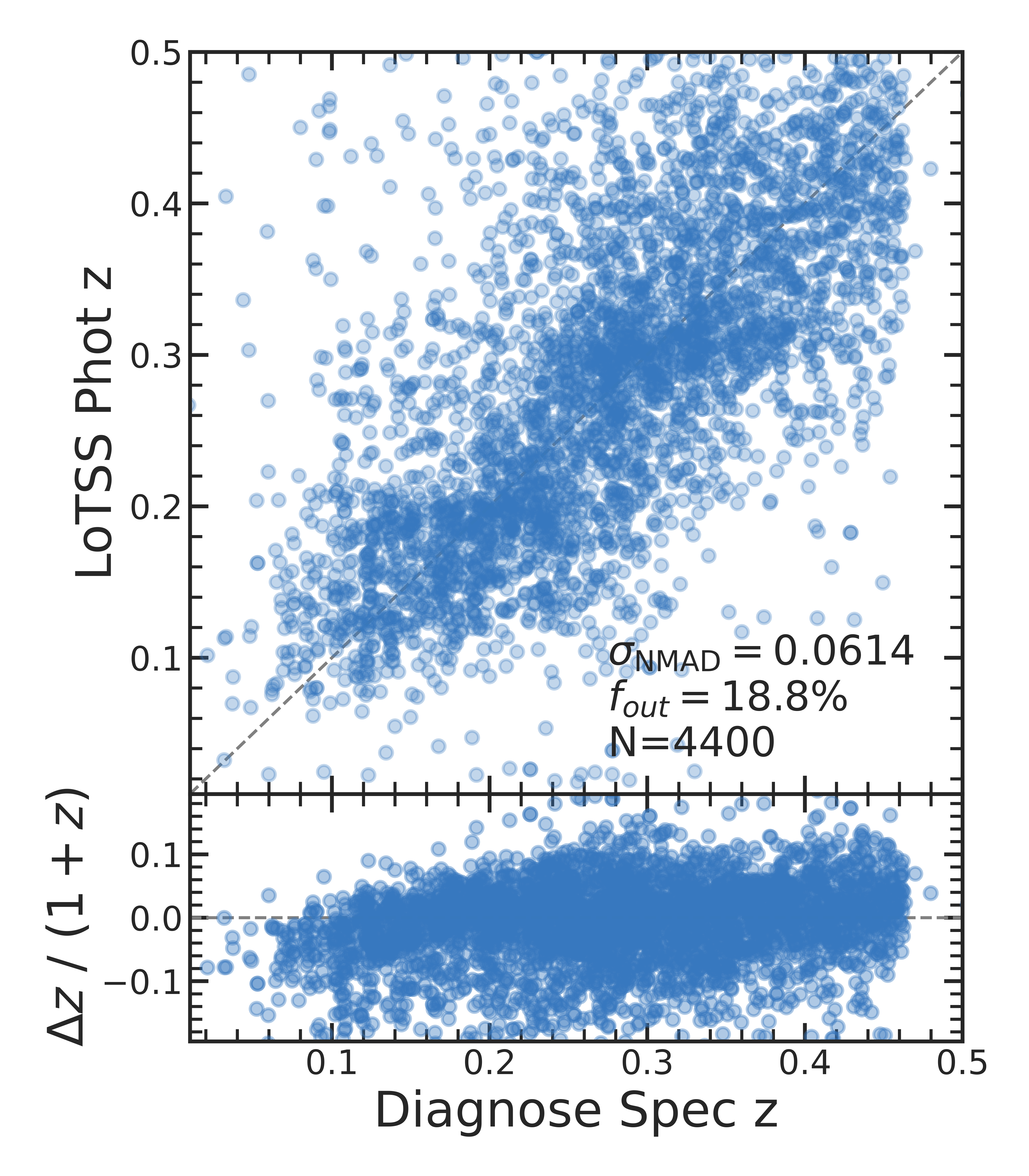}
  \caption{Comparison between LoTSS photometric redshifts and Diagnose spectroscopic redshifts for the same sources. In total, there were 4,400 LoTSS sources with photometric redshifts and no previous spectroscopic redshift counterparts. The vertical bound created by the data points is a result of the cutoff at highest possible redshift at VIRUS wavelengths for an \OII\ emitter. The normalized median absolute deviation (NMAD) and outlier fraction are calculated as in \cite{momcheva16}. This relationship is to be expected for a photo-\textit{z} to spec-\textit{z} comparison.}
  \label{fig:knownphotz}
\end{figure}

We also compared our spectroscopic redshifts to the photometric estimates from the LoTSS value-added catalog. Figure~\ref{fig:knownphotz} shows this comparison over the range $0.0 < z < 0.5$; we find a good agreement between the two measurements with \textrm{${\sigma}_z = 0.0614$}.

\subsection{Classification and Redshift Methodology}

\begin{figure}[t]
  \centering
  \includegraphics[width=\columnwidth]{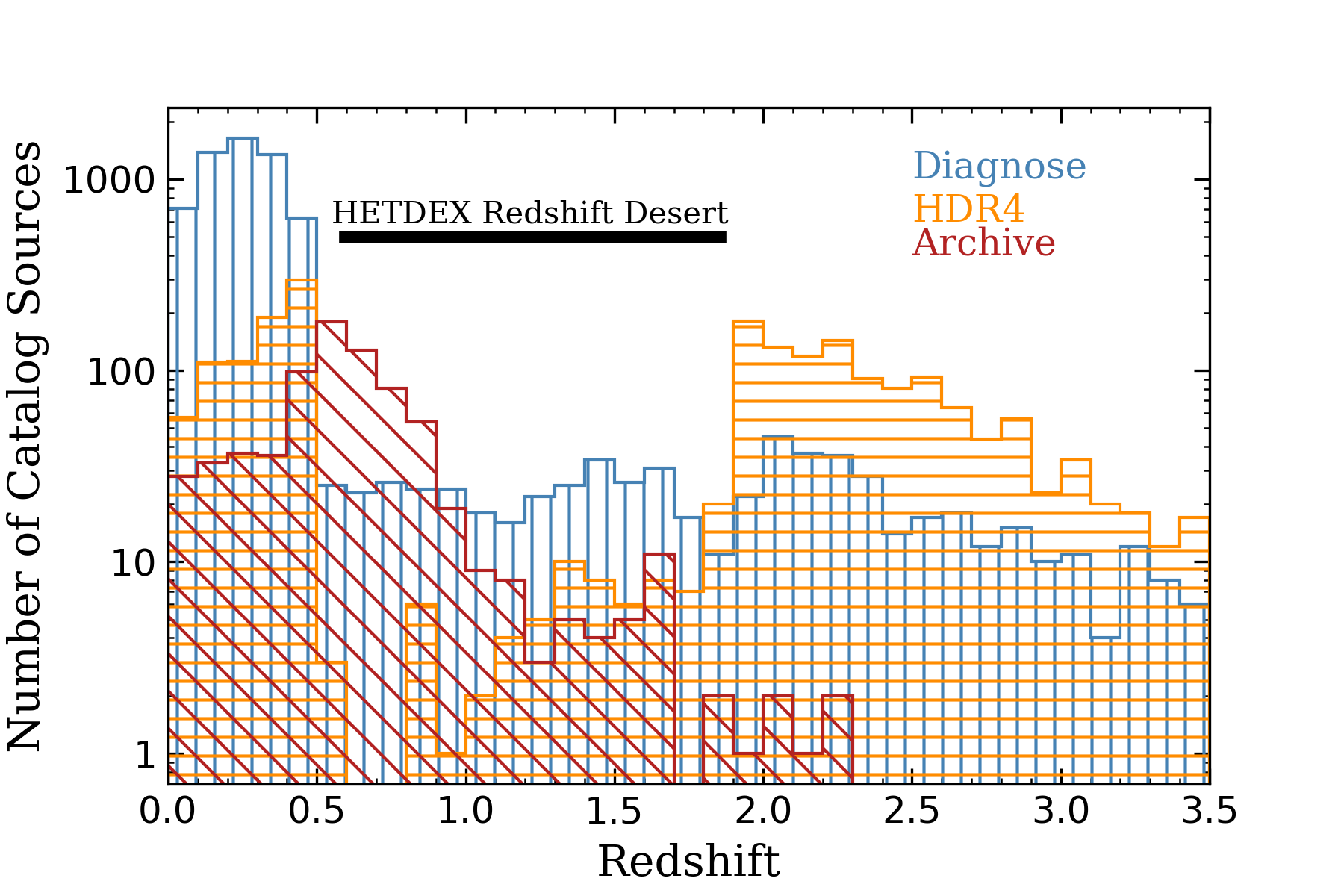}
  \caption{We show the distribution of redshifts in the HETDEX-LOFAR spectroscopic catalog. There are \numredshifts\ sources in the catalog, the majority of which are between $0.0 < z < 0.5$ and $1.9 < z < 3.5$. There is a void of redshifts between those two regions due to a lack of strong features present in the VIRUS wavelength bandpass. The stacked histogram in blue shows the redshifts from Diagnose, which make up the majority of the sample, are mostly lower redshift objects. The stacked histogram in orange shows the redshifts from HDR4; these are more evenly distributed between low and high redshift. Finally, the stacked histogram in red are the redshifts from the archive not in the other two classifiers.}
  \label{fig:redshiftdist}
\end{figure}

\begin{figure*}[hbt!]
  \centering
  \includegraphics[width=2\columnwidth]{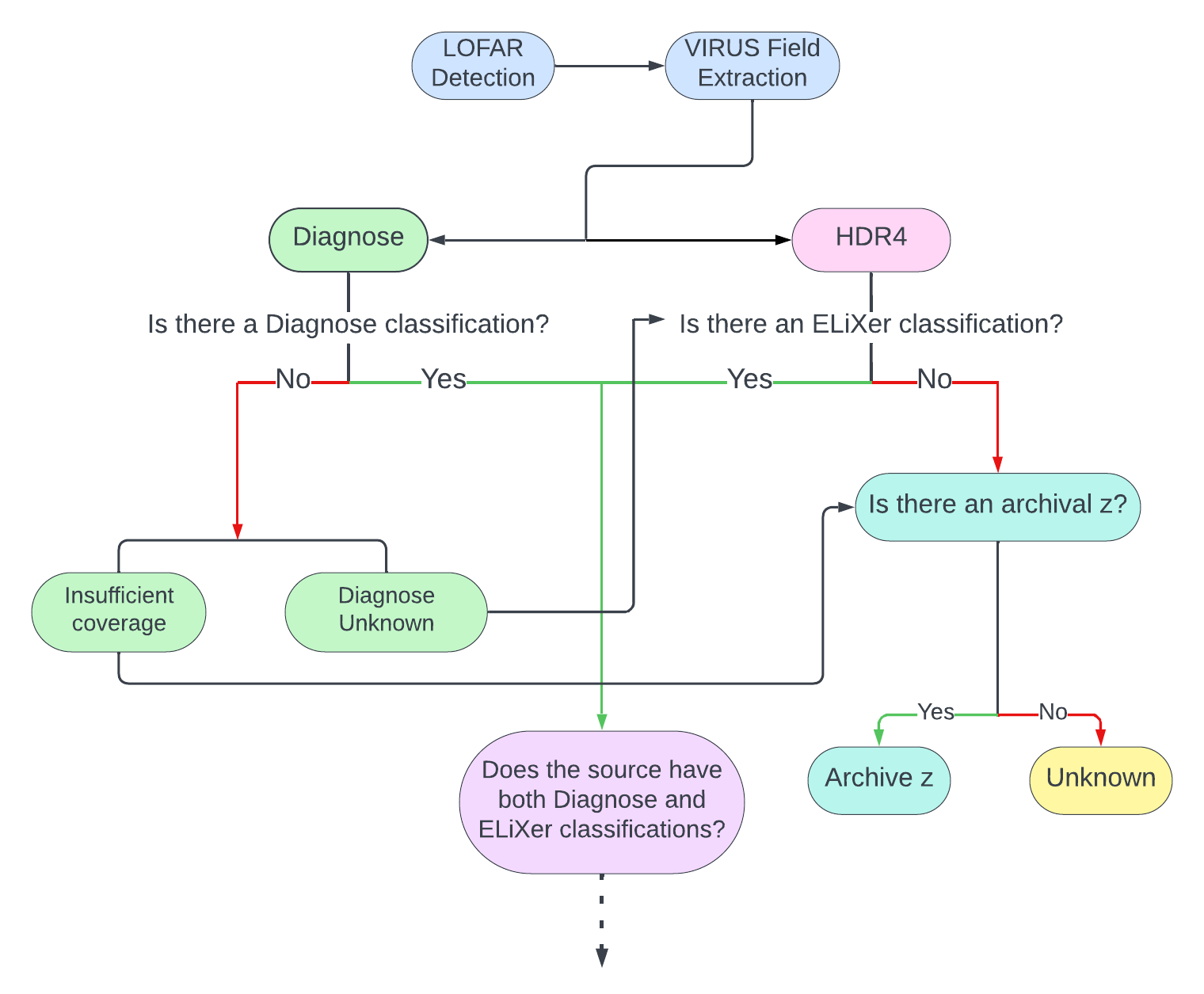}
  \caption{HETDEX-LOFAR Redshift and Classification Pipeline. The process begins with the LOFAR detection, followed by the extraction of HETDEX spectra. Once we have the HETDEX spectra, we run them through Diagnose and ELiXer to determine whether they have a classification and redshift from either source. If they have either a Diagnose or ELiXer redshift, we accept this value; if they have both and the redshifts disagree, we perform a detailed examination for the reason behind the discrepancy.
  The specifics of this examination are detailed in Appendix \ref{appendix:a}. If neither Diagnose nor ELiXer provide a redshift, we see if there is an archival value. If so, we accept this value. If not, the source remains unclassified. }
  \label{fig:detection_pipeline}
\end{figure*}

The HETDEX-LOFAR spectroscopic redshifts and classifications have three origins: Diagnose, HDR4, and archival. We discussed the combination of Diagnose and HDR4 redshifts in \S\ref{sec:combine}, and with all three origins we follow a similar logic. If Diagnose has a robust redshift, this is used. If not and HDR4 has a redshift, then this is used. Finally, if neither Diagnose nor HDR4 provide a redshift for the source, but the value-added catalog does then we use the archival spectroscopic redshift. We prioritize archival redshifts last in order to maximize the number of new redshifts determined for this catalog. Figure~\ref{fig:redshiftdist} shows the final redshift distribution for the \numredshifts\ sources and the origin of the redshift. The overall detection and classification pipeline is outlined in Figure~\ref{fig:detection_pipeline}. 

Our classification follows mostly from the redshift of the source and the origin catalog. We group all sources labeled `STAR' by Diagnose or HDR4 together as `STAR'. We group all objects labeled `QSO' or `AGN' in Diagnose or HDR4, respectively, as `AGN'; this is done for all redshifts $0.0 < z < 3.5$. Classifications of `LZG' and `OII' from HDR4 and `GALAXY' from Diagnose are all grouped under the label `LOWZGAL' for galaxies $0.0 < z < 0.5$. Classifications of `LAE' from HDR4 are grouped as `HIGHZGAL' for systems with $1.9 < z < 3.5$. Finally, if the redshift comes from the archive, we label the group `ARCHIVE' with $0.0 < z < 3.5$. So our final five labels are `STAR', `AGN', `LOWZGAL', `HIGHZGAL', and `ARCHIVE'. Table~\ref{table:breakdown} includes a breakdown of the number of sources at different steps of the classification process, including the final catalog size.

\begin{table*}[t]
 
\centering
\begin{tabular}{||c l ||} 
 \hline
 Number of Sources & Description   \\ [0.5ex] 
 \hline\hline
 325,694 & Sources in LoTSS DR1  \\
 28,705 & Spectral matches between LoTSS DR1 \& HETDEX DR4   \\
 4,908 & Sources w/ Diagnose \& ELiXer redshifts  \\
 9,710 &  \# of spectroscopic redshifts in final HETDEX-LOFAR catalog \\
 9,087 & New spectroscopic redshifts in final catalog  \\
 197 &  `STAR' in final catalog \\
 804 & `AGN' in final catalog   \\
 6,394 & `LOWZGAL' in final catalog  \\
 1,075 & `HIGHZGAL' in final catalog   \\
 757 & `ARCHIVE' in final catalog   \\
 \hline
\end{tabular}
  \caption{Breakdown of the number of sources relevant to different steps of the classification and redshift assignment pipeline including final spectroscopic redshift catalog size.}
\label{table:breakdown}
\end{table*}

\subsection{HETDEX-LOFAR Spectroscopic Catalog}

\begin{figure*}[t]
  \centering
  \includegraphics[width=2\columnwidth]{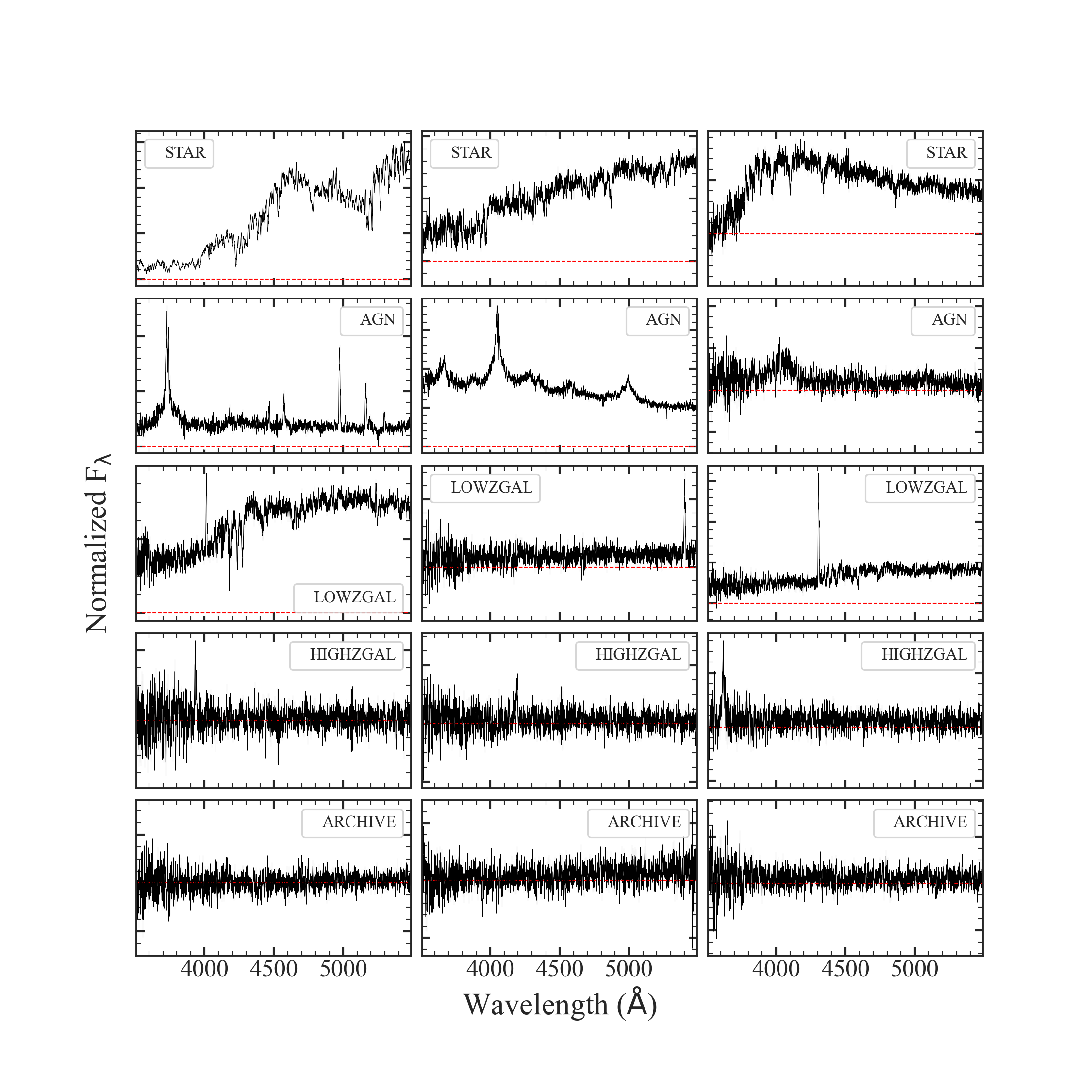}
  \caption{We show 15 example spectra from the HETDEX extractions of LoTSS sky positions: three each with labels of `STAR', `AGN', `LOWZGAL', `HIGHZGAL', and `ARCHIVE.' We also mark the null flux density with a dashed red line.}
  \label{fig:examplespeclabels}
\end{figure*}

Our final compiled spectroscopic redshift catalog includes \numredshifts\ total redshifts: \nstars\ `STAR', \nagns\ `AGN', \nlowzgal\ `LOWZGAL', \nhighzgal\ `HIGHZGAL', and \narchive\ `ARCHIVE'. Table~\ref{table:ascii} explains the column names for the final catalog. In Figure~\ref{fig:examplespeclabels}, we show three example spectra for each of the five labels.

\begin{table*}[t]
 
\centering
\begin{tabular}{||>{\columncolor[gray]{0.8}}c l c ||} 
 \hline
 Column Name & Description  & Data Type \\ [0.5ex] 
 \hline\hline
 objID & LoTSS Object ID  & string \\
 source\_name & Source Name  & string \\
 RA & PanSTARRS1 Right Ascension (J2000) &  float \\
 Dec & PanSTARRS1 Declination (J2000) & float\\
 z\_diagnose & Best fit redshift from Diagnose  &float \\
 z\_hdr4 & Best fit redshift from ELiXer & float \\
 z\_archive & Spectroscopic redshift from value-added LoTSS catalog  &float \\
 z\_best & HETDEX-LOFAR redshift  &float \\
 z\_best\_src & 1 $=$ Diagnose, 2 $=$ HDR4, 3 $=$ Archive  & integer \\
 classification & STAR, AGN, LOWZGAL, HIGHZGAL, or ARCHIVE  &string \\
 log\_mass & \texttt{MCSED} derived stellar mass  & float \\
 log\_SFR & \texttt{MCSED} derived star formation rate & float \\
 log\_L150 & \texttt{MCSED} derived 150 MHz luminosity & float\\
 \hline
\end{tabular}
  \caption{Description of the catalog columns. `z best' and `classification' reflect final HETDEX-LOFAR catalog values.}
\label{table:ascii}
\end{table*}

\section{Star Formation at Radio Wavelengths} 
\label{sec:discussion}

\subsection{Spectral Energy Distribution Fitting}

We limited our analysis of HETDEX-LOFAR galaxies to the 6,499 objects with $0.01 < z < 0.47$ to ensure \OII\ was in the VIRUS bandpass. For each galaxy, we collected the Pan-STARRS $grizy_{\rm P1}$ and WISE W1W2 photometry in the \citet{Williams19} catalog, and then further restricted our sample to those sources with a Pan-STARRS $g_{\rm P1}$-band detection; this reduced our sample to 5,919 systems. Although this photometry alone is often enough to derive quantities such as SFR and stellar mass from spectral energy distributions, we can also utilize our VIRUS spectroscopy to further inform the fitting.  For consistency, we normalized our VIRUS spectra to the Pan-STARRS $g_{\rm P1}$-band photometry, and then calculated 10 synthetic narrowband values in chunks of 200\,\AA\ across the bandpass. We also used \texttt{ppxf} \citep{Cappellari23} to quickly model the underlying stellar continuum to measure the \OII\ emission. Using the 7 bands of photometry, 10 synthetic narrow bands of the VIRUS spectroscopy, and \OII\ emission, we estimated the stellar masses, star formation rates, and dust attenuation via SED fitting using \texttt{MCSED} \citep{Bowman20}.

\texttt{MCSED} is a flexible SED fitting code that allows users to supply both photometry and emission line fluxes to fit the stellar populations of a galaxy. \texttt{MCSED} implements a stellar library generated by the Flexible Stellar Population Synthesis (FSPS) code \citep{Conroy09,Conroy10} employing PADOVA isochrones \citep{Bertelli94,Girardi00,Marigo08}, a self-consistent prescription for nebular line and continuum emission given by the grid of CLOUDY models \citep{Ferland98,Ferland13} generated by \citet{Byler17}, and a \citet{Chabrier03} initial mass function (IMF\null). We adopted an eleven-parameter model, with the variables being stellar metallicity (ranging from $\sim$1\% to 150\% solar metallicity), a non-parametric six-age-bin star formation history (using a constant SFR within each bin (defined at ages of 0.001, 0.03, 0.1, 0.3, 1.0, 3.6, 13.2 Gyr), a single parameter dust attenuation law \citep{Calzetti00}, and a three-parameter dust emission model from \citet{Draine-Li07} constrained by energy balance between absorption and emission. The nebular metallicity was fixed to the stellar metallicity and the ionization parameter of the nebular emission was fixed at Log(U) = -2.5. Changing the adopted fitting assumptions (especially the SFR history) can systematically affect the stellar masses at the level of $\sim$0.3 dex \citep{Conroy09,Conroy10}. 

\texttt{MCSED} utilizes the \texttt{emcee} Python module \citep{Foreman13} with initial positions defined by a random Gaussian ball near the middle of the range of allowed values for each parameter and a small but generous sigma to avoid the initial boundaries, yet explore the available phase space. We ran the fitting using a Monte Carlo Markov Chain (MCMC) approach with 40 walkers and 800 steps. Convergence is always a challenge in Monte Carlo methods, and with 11 free parameters, the choice of 40 walkers and 800 steps was a compromise between convergence and computation cost. 

\subsection{Star Formation Rate and 150 MHz Luminosity}

\begin{figure*}[t]
  \centering
  \includegraphics[width=2\columnwidth]{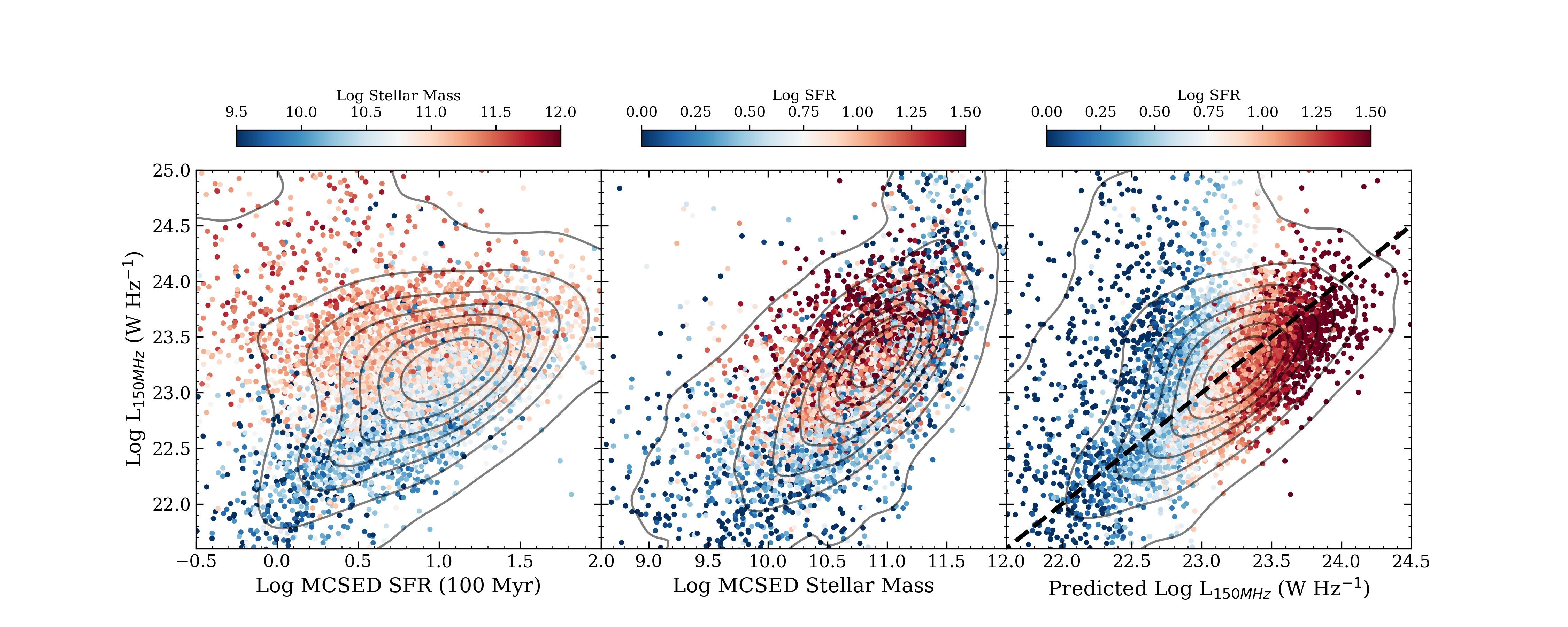}
  \caption{MCSED results for the 6,499 galaxy sample with $0.01 < z < 0.47$. The left panel demonstrates a tight correlation between SFR and 150 MHz luminosity with expected scatter from the secondary mass dependence show in \citet{Smith21}. The individual points are colored by stellar mass and show this dependence. The center panel demonstrates the correlation between stellar mass and 150 MHz luminosity, as colored by SFR\null. The right plot shows the correlation between our derived 150 MHz luminosity and the \citet{Smith21} predicted 150 MHz luminosity. The black dashed line represents a one-to-one correlation. The individual points are colored by SFR.}
  \label{fig:hdr4_mcsed}
\end{figure*}

For our sample of galaxies, we used the \texttt{MCSED} results to examine the correlation between 150 MHz luminosity and SFR, as well as the secondary stellar mass dependence. Figure~\ref{fig:hdr4_mcsed} shows the results for each of these relationships. All of the galaxies studied have SFR and stellar mass estimates that were derived from energy balance spectral energy distribution fitting using redshifts and aperture-matched forced photometry from the LOFAR Two-metre Sky Survey (LoTSS) Deep Fields data release. The first panel in Figure~\ref{fig:hdr4_mcsed} shows the correlation between SFR and 150 MHz luminosity. There is tight correlation between the two quantities, though there is some scatter caused by the secondary mass dependence acknowledged by \citet{Smith21}. This mass dependence can be seen via the color bar. The second panel demonstrates the correlation between stellar mass and 150 MHz luminosity. There is a strong correlation here, which is to be expected as we anticipate a secondary mass dependence. Additionally, we referenced the \citet{Smith21} calculation for 150 MHz luminosity to determine the predicted luminosity expected for our \texttt{MCSED} results which is shown in the third panel. The black dashed line represents a one-to-one correlation. All of our data follows this trend and is tightly correlated. This comparison acts as a check on our derived values for 150 MHz luminosity, stellar mass, and SFR. The majority of our derived values match closely to those predicted by the Smith relation. Overall, we were able to measure SFR, 150 MHz luminosity, and stellar mass for 6,499 galaxies.

\subsection{Line Ratio Diagnostics}

We used \texttt{PPXF} to measure the line fluxes for our individual galaxies, particularly focusing on the \OIII, \OII, and \NeIII\ emission lines. We measure these lines in particular due to their use in radio astronomy. \OIII\ is used to trace ionized outflows from radio sources because of its sensitivity to the impact of radiation and jets \citep[\eg][]{kukreti23}. Additionally, these three lines can be useful in line ratios as indicators of ionization parameter. The most commonly used diagnostic of the ionization parameter is \OIII/\OII\ (O3O2) \citep[\eg][]{alloin78,baldwin81}; however, the wavelength range between \OIII\ and \OII\ makes this line ratio diagnostic radio sensitive to extinction effects. As an alternative, \NeIII/\OII\ (Ne3O2) can act as a similar diagnostic of ionization parameter that is radio insensitive to reddening effects \citep[\eg][]{levesque14}. The similar short wavelengths of \NeIII\ and \OII\ also make the diagnostic usable at larger redshifts (\textit{z} $\sim$ 1.6) than O3O2 \citep{nagao06}. The benefits of Ne3O2 as a diagnostic of ionization parameter as compared to those of O3O2 led us to solely examine the Ne3O2 line ratio in our sample. Because many of the measured lines are weak, we also computed the biweight stack of the HETDEX spectra binned by stellar mass, with each spectrum normalized by its median continuum value in the rest-frame wavelength range of $3750\, {\rm \AA} < \lambda < 3850$\,\AA\ for Ne3O2.  We also limited our sample to galaxies with $z < 0.4$ to ensure we detect the \NeIII\ and \OII\ lines.

The line ratio diagnostics used are indicators of ionization and relationship to AGN activity; however, they are not definitive criteria to determine contribution from AGN activity. To further understand the properties of our data set, we used the relation between SFR and 150 MHz luminosity derived in \citet{Best23}. This relation can be used to set an N-$\sigma$ cutoff above the \citet{Best23} ridgeline which can help determine which galaxies are star-forming and which are dominated by AGN activity. Galaxies that fall within the N-$\sigma$ cutoff are star-forming, while those above the cutoff have AGN activity.

\begin{figure}[t]
  \centering
  \includegraphics[width=\columnwidth]{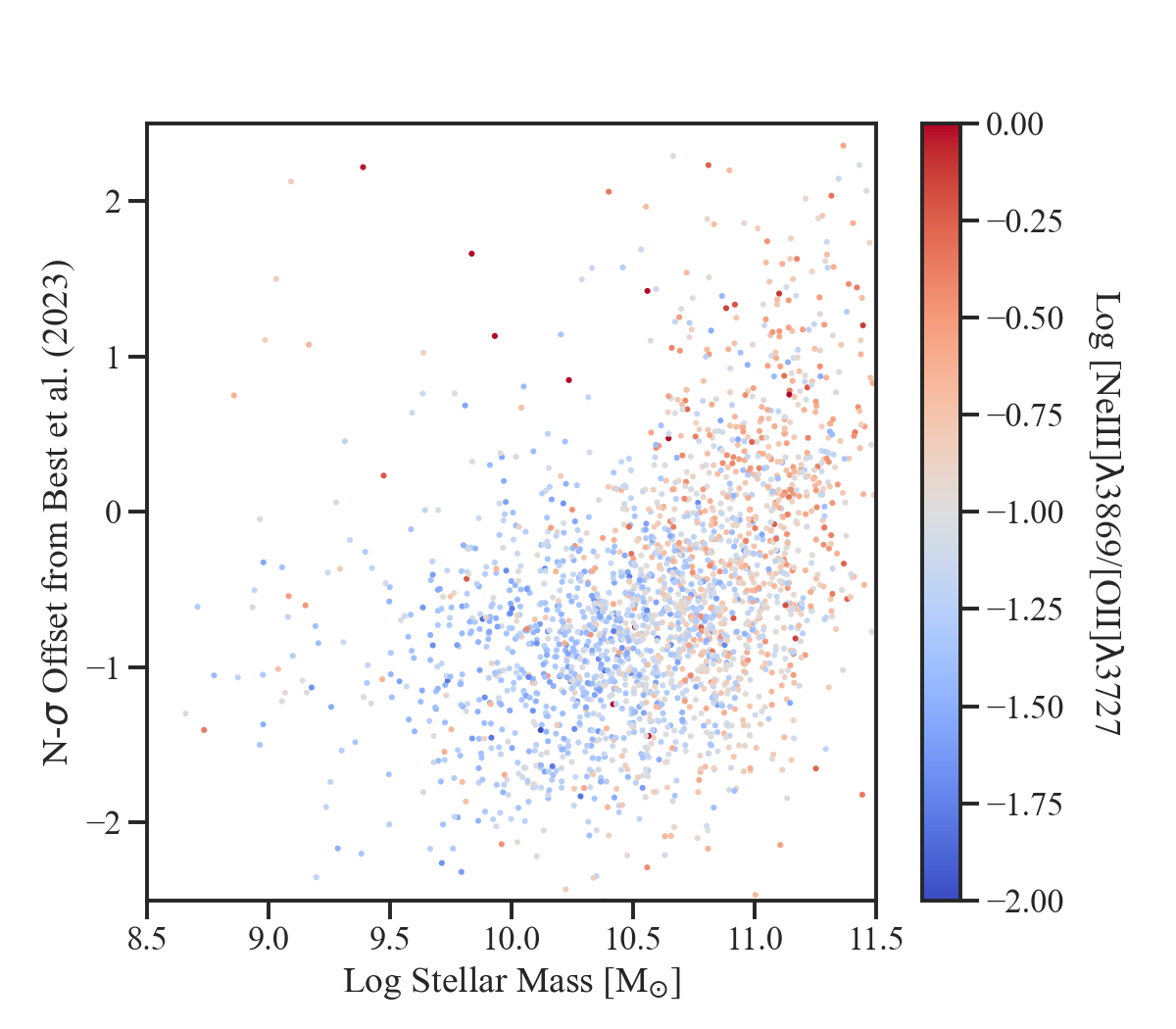}
  \caption{Stellar mass vs. N-$\sigma$ offset from the \citet{Best23} relation between SFR and 150 MHz luminosity colored by log$_{10}$(Ne3O2) line ratio. AGN are anticipated to have log$_{10}$(Ne3O2) $> 1$, but none of our sources exhibit such ionization hardness. At $\log_{10}(M/M_{\odot}) > 10.5$, there is an increase in offset with Ne3O2, indicating that there could be more AGN contribution for these sources.}
  \label{fig:massoffset}
\end{figure}

We further explored this relation by examining the relationship between stellar mass and the N-$\sigma$ offset from the \citet{Best23} relation (Figure~\ref{fig:massoffset}). AGN typically have harder ionization fields, meaning that the their log$_{10}$(Ne3O2) line ratios should be greater than 1. By coloring the scatter in Figure~\ref{fig:massoffset} by the Ne3O2 ratio, we see that none of the galaxies with $z < 0.4$ have log$_{10}$(Ne3O2) $> 1$; however, as galaxies reach $\log_{10}(M/M_{\odot}) > 10.5$, Ne302 increases with the offset from the \citet{Best23} relationship. This suggests that there could be a greater contribution from AGN\null.

\subsection{The SFR-150 MHz Luminosity Relation}

We determined the \sfrL\ relation as follows. First, we applied a mass cut at log$_{10}(M/M_\odot) < 11.0$. This mass cut arises from the analysis of Figure~\ref{fig:massoffset}, which shows increasing AGN contribution above log$_{10}$$(M/M_\odot) \approx 10.5$. To ensure that our fit was based mostly on star-forming galaxies, we removed the population with potential AGN contribution from our sample used for fitting. For the remaining sources, we calculated the errors on  SFR, stellar mass, and L$_{\rm{150MHz}}$ and found an average log$_{10}$ error of $\pm 0.216$ [$M_\odot yr^{-1}$] for SFR, $\pm 0.132$ [$M_\odot$] for stellar mass, and $\pm 0.070$ [W Hz$^{-1}$] for L$_{\rm{150MHz}}$.

To determine best-fit parameters, we adopted the form of the mass-dependent \sfrL\ relationship from \cite{gurkan18}:
\begin{equation}
L_{\rm{150MHz}}=L_C\psi^\beta(\frac{M_*}{10^{10}M_\odot})^\gamma
\end{equation}
where L$_C$ is the 150 MHz luminosity of a galaxy with $M_*=10^{10}M_\odot$ and $\psi=1$ $M_\odot yr^{-1}$. We find the best-fit values of log$_{10}$$L_C=22.341\pm0.016$, $\beta=0.526\pm0.017$, and $\gamma=0.384\pm0.017$, determined using the \texttt{emcee} \citep{dfm19} Monte Carlo Markov Chain (MCMC) algorithm with 15 walkers and a chain length of 10,000 samples. Our best fit \sfrL\ relation is, therefore, log$_{10}L_{\rm{150MHz}}$ = (22.341$\pm$0.016) + (0.526$\pm$0.017) log$_{10}$($\psi$/$M_\odot yr^{-1}$) + (0.384$\pm$0.017) log$_{10}$($M/10^{10}M_\odot$). 

\begin{figure}[t]
    \centering
    \includegraphics[width=\columnwidth]{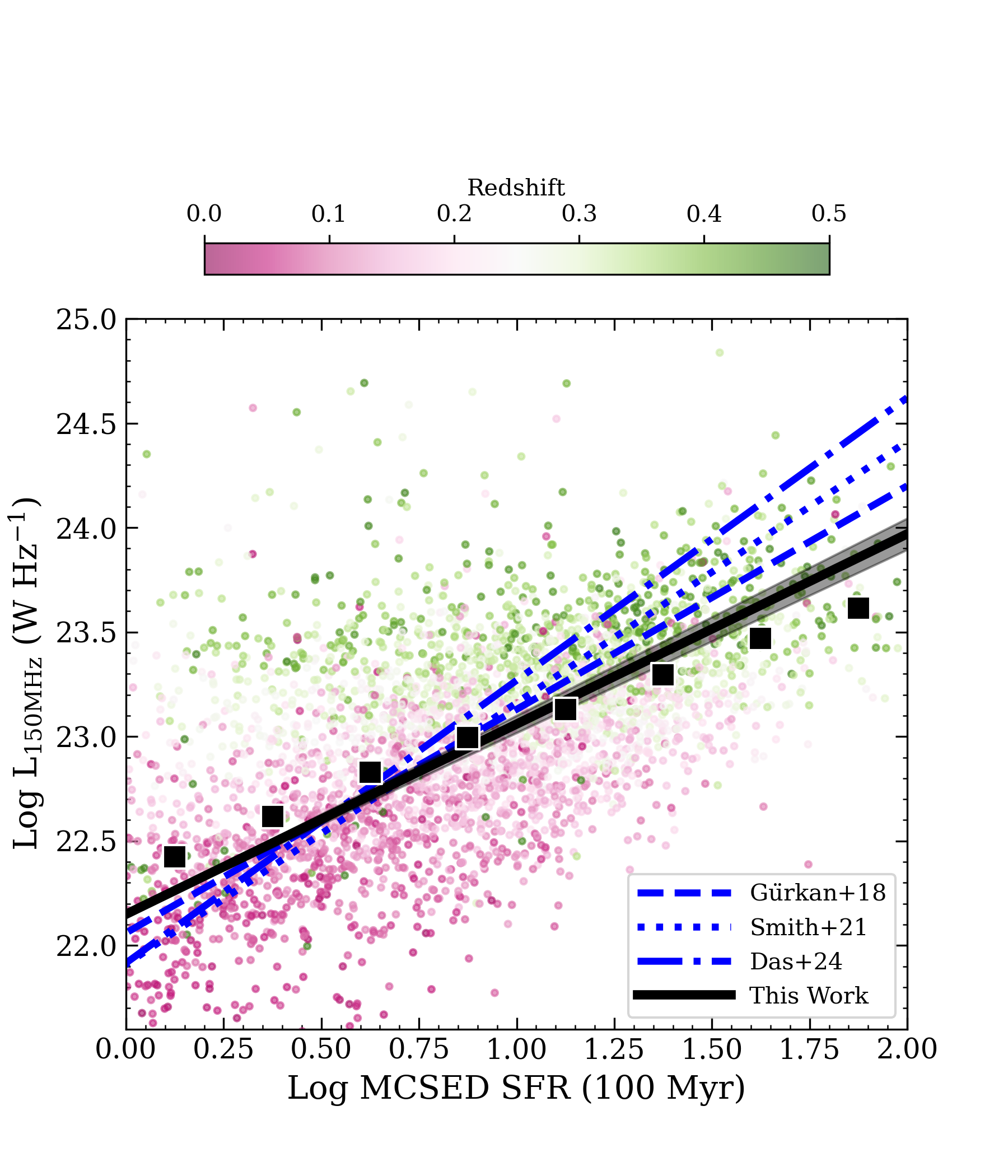}
    \caption{Distribution of derived radio luminosity versus MCSED star formation rate (100 Myr) for the HETDEX-LOFAR galaxies with 0.01 $<$ \textit{z} $<$ 0.47. The points represent the portion of the sample with log$_{10}$($M/M_\odot$) $<$ 11.0 and are colored by redshift. The black squares represent the biweight of log$_{10}$(L$_{\rm{150MHz}}$) for bins of log$_{10}(\psi)$. Our derived relationship for radio luminosity versus star formation rate is shown by the black solid line. Different fits from literature are shown in with blue lines (dashed - \cite{gurkan18}; dotted - \cite{Smith21}; dashed and dotted - \cite{das24}).}
    \label{fig:hxlf_relation}
\end{figure}

We compared this mass-dependent line fit to those found by \cite{gurkan18}, \cite{Smith21}, and \cite{das24} in Figure~\ref{fig:hxlf_relation}. \cite{gurkan18} obtained best-fit estimates of log$_{10}$$L_C=22.13\pm0.01$, $\beta=0.77\pm0.01$, and $\gamma=0.43\pm0.01$. \cite{Smith21} found best-fit estimates of log$_{10}$$L_C=22.218\pm0.016$, $\beta=0.903\pm0.012$, and $\gamma=0.332\pm0.037$. \cite{das24} obtained best-fit estimates of log$_{10}$$L_C=22.083\pm0.004$, $\beta=0.778\pm0.004$, and $\gamma=0.334\pm0.006$. Our line has a shallower slope than the three previously derived mass-dependent expressions. This difference could be a result of several factors. Our relation is derived using spectroscopic redshifts with small error, creating an upper radio luminosity limit that is quite sharp (shown by the coloring of data points in Figure~\ref{fig:hxlf_relation}). This upper limit could be depressing the steepness of our fit because it is creating an upper limit on radio luminosity. There is also a lower limit caused by the flux density limit of our sample. The combination of both the upper and lower limits on luminosity could result in a lack of objects that would tend to populate the lower-left and upper-right of Figure~\ref{fig:hxlf_relation}. The slope of the fit is most influenced by objects at the extremes of radio luminosity and SFR, so if there is limit-related bias present, the slope will also be biased. Additionally, the applied mass cut could have removed a number of the high luminosity sources, which would also create bias in the slope. The larger sample that will be available with the release of the full
HETDEX survey -- approximately 40,000 sources -- will allow a more thorough analysis of the possible systematic effects in fitting the SFR relation, and this analysis can be investigated further in a future paper.

\section{Summary}
\label{sec:summary}
Combining data from an optical spectroscopic survey, the fourth data release of the Hobby-Eberly Telescope Dark Energy Experiment (HETDEX) catalog and a radio survey, the first data release of the Low Frequency Array Two-metre Sky Survey (LoTSS), we were able to determine intrinsic properties for radio sources present in both fields of view.  Using positions in LoTSS we extracted 18,267 spectra from the HETDEX database. We used a robust and automatic classification code called Diagnose, developed by our group for the HET VIRUS Parallel Survey, to determine redshifts and object classifications from the optical spectra. We also use the HETDEX ElixerWidget to source redshifts and object classifications.  To determine redshifts and classifications for the remaining unknown spectra, we matched these sources (2\arcsec\ radius) to object positions in the HETDEX data release 4 catalog. Using these methods, we created the HETDEX-LOFAR Spectroscopic Redshift Catalog with \numredshifts\ total redshift values. We group all sources labeled `STAR' by Diagnose or HDR4 together as `STAR'. We group all objects labeled `QSO' or `AGN' in Diagnose or HDR4, respectively, as `AGN'; this is done for all redshifts $0.0 < z < 3.5$. Classifications of `LZG' and `OII' from HDR4 and `GALAXY' from Diagnose are all grouped under the label `LOWZGAL' for galaxies $0.0 < z < 0.5$. Classifications of `LAE' from HDR4 are grouped as `HIGHZGAL' for systems with $1.9 < z < 3.5$. Finally, if the redshift comes from the archive, we label the group `ARCHIVE' with $0.0 < z < 3.5$. So our final five labels are `STAR', `AGN', `LOWZGAL', `HIGHZGAL', and `ARCHIVE'.

The compiled catalog includes \nstars\ `STAR', \nagns\ `AGN', \nlowzgal\ `LOWZGAL', \nhighzgal\ `HIGHZGAL', and \narchive\ `ARCHIVE' sources. 

The focus of this project is assigning redshifts for sources, which, for extragalactic objects, allows one to determine distances and hence many intrinsic properties, most importantly the luminosity. Using line ratio diagnostics such as \NeIII/\OII, we probed the ionization parameter of the gas. These properties enable AGN excitation to be detected within the star-forming galaxies.

The HETDEX-LOFAR Spectroscopic Redshift Catalog contains the highest substantial fraction of LOFAR galaxies with spectroscopic redshift information and coverage. The catalog also offers an improvement over the archival spectroscopic redshift estimates provided in the LoTSS value-added catalog. It also enables the investigation of SFR tracers with high quality data, though the results may not be definitive. We derive the SFR, stellar masses, 150 MHz luminosity, and emission lines for $\sim 75\%$ of our sources with  $z < 0.4$, as well as fit a new \sfrL\ relationship. Understanding the relationship between radio luminosities and SFR is increasingly important in the upcoming era of SKA, and this work acts as the first step to help inform future work in the radio community such as investigating the connection between ionized outflows traced by \OIII\ and radio emission, refining radio luminosity functions, and comparison of spectra with resolved sub-galactic radio emission. All of the values derived through this work can act as a reference point for adjusting star formation surveys.

The HETDEX-LOFAR Spectroscopic Redshift Catalog provides key physical properties of \numredshifts\ including RA, Dec, spectroscopic redshift, classification, stellar mass, SFR, and 150 MHz luminosity. These properties will serve to enable science investigations in the radio astronomy community.

\subsection{Data Release}

The fourth internal data release of HETDEX covers $\sim 90\%$ of the total survey, making this the first paper in a series about combining HETDEX and LoTSS. By the time HETDEX finishes, we anticipate a final HETDEX-LOFAR sample of $\sim$ 40,000 galaxies. This paper includes the release of derived spectroscopic redshifts, classifications, stellar masses, SFR, 150 MHz luminosity, and line fluxes for each source, as well as the spectra. A copy of the HETDEX-LOFAR Spectroscopic Redshift Catalog is available on Zenodo doi:\url{10.5281/zenodo.13619775}. This Zenodo deposit includes a FITS file with the spectra for all 28,705 sources, as well as the derived redshifts, classifications, and \texttt{MCSED} quantities for each source (columns described in Table~\ref{table:ascii}). The deposit also includes the required statements and papers to reference to acknowledge use of data from the HETDEX survey.

\section*{Acknowledgments}
HETDEX is led by the University of Texas at Austin McDonald Observatory and Department of Astronomy with participation from the Ludwig-Maximilians-Universit{\" a}t M{\" u}nchen, Max-Planck-Institut f{\" u}r Extraterrestrische Physik (MPE), Leibniz-Institut f{\" u}r Astrophysik Potsdam (AIP), Texas A\&M University, Pennsylvania State University, Institut f{\" u}r Astrophysik G{\" o}ttingen, The University of Oxford, Max-Planck-Institut f{\" u}r Astrophysik (MPA), The University of Tokyo and Missouri University of Science and Technology.

Observations for HETDEX were obtained with the Hobby-Eberly Telescope (HET), which is a joint project of the University of Texas at Austin, the Pennsylvania State University, Ludwig-Maximillians-Universit{\" a}t M{\" u}nchen, and Georg-August-Universit{\" a}t, G{\" o}ttingen. The HET is named in honor of its principal benefactors, William P. Hobby and Robert E. Eberly. We thank the staff at McDonald Observatory for making this project possible. The Visible Integral-field Replicable Unit Spectrograph (VIRUS) was used for HETDEX observations. VIRUS is a joint project of the University of Texas at Austin, Leibniz-Institut f{\" u}r Astrophysik Potsdam (AIP), Texas A\&M University, Max-Planck-Institut f{\" u}r Extraterrestrische Physik (MPE), Ludwig-Maximilians-Universit{\" a}t M{\" u}nchen, Pennsylvania State University, Institut f{\" u}r Astrophysik G{\" o}ttingen, University of Oxford, and the Max-Planck-Institut f{\" u}r Astrophysik (MPA).

Funding for HETDEX has been provided by the partner institutions, the National Science Foundation, the State of Texas, the US Air Force, and by generous support from private individuals and foundations.

The authors acknowledge the Texas Advanced Computing Center (TACC) at The University of Texas at Austin for providing computing resources that have contributed to the research results reported within this paper. URL:\url{ http://www.tacc.utexas.edu}.

The Institute for Gravitation and the Cosmos is supported by the Eberly College of Science and the Office of the Senior Vice President for Research at the Pennsylvania State University.

The Pan-STARRS1 Surveys (PS1) and the PS1 public science archive have been made possible through contributions by the Institute for Astronomy, the University of Hawaii, the Pan-STARRS Project Office, the Max-Planck Society and its participating institutes, the Max Planck Institute for Astronomy, Heidelberg and the Max Planck Institute for Extraterrestrial Physics, Garching, The Johns Hopkins University, Durham University, the University of Edinburgh, the Queen's University Belfast, the Harvard-Smithsonian Center for Astrophysics, the Las Cumbres Observatory Global Telescope Network Incorporated, the National Central University of Taiwan, the Space Telescope Science Institute, the National Aeronautics and Space Administration under Grant No. NNX08AR22G issued through the Planetary Science Division of the NASA Science Mission Directorate, the National Science Foundation Grant No. AST-1238877, the University of Maryland, Eotvos Lorand University (ELTE), the Los Alamos National Laboratory, and the Gordon and Betty Moore Foundation.

We also thank the Erickson Discovery Grant for providing support to complete this project during Summer 2022. Funding from this grant paved the path to complete the majority of analysis required for this project.

\bibliography{bibliography}{}
\bibliographystyle{aasjournal}

\appendix
\twocolumngrid
\section{Diagnose and HDR4 Redshift}
\label{appendix:a}

\begin{figure*}[]
  \centering
  \includegraphics[width=2\columnwidth]{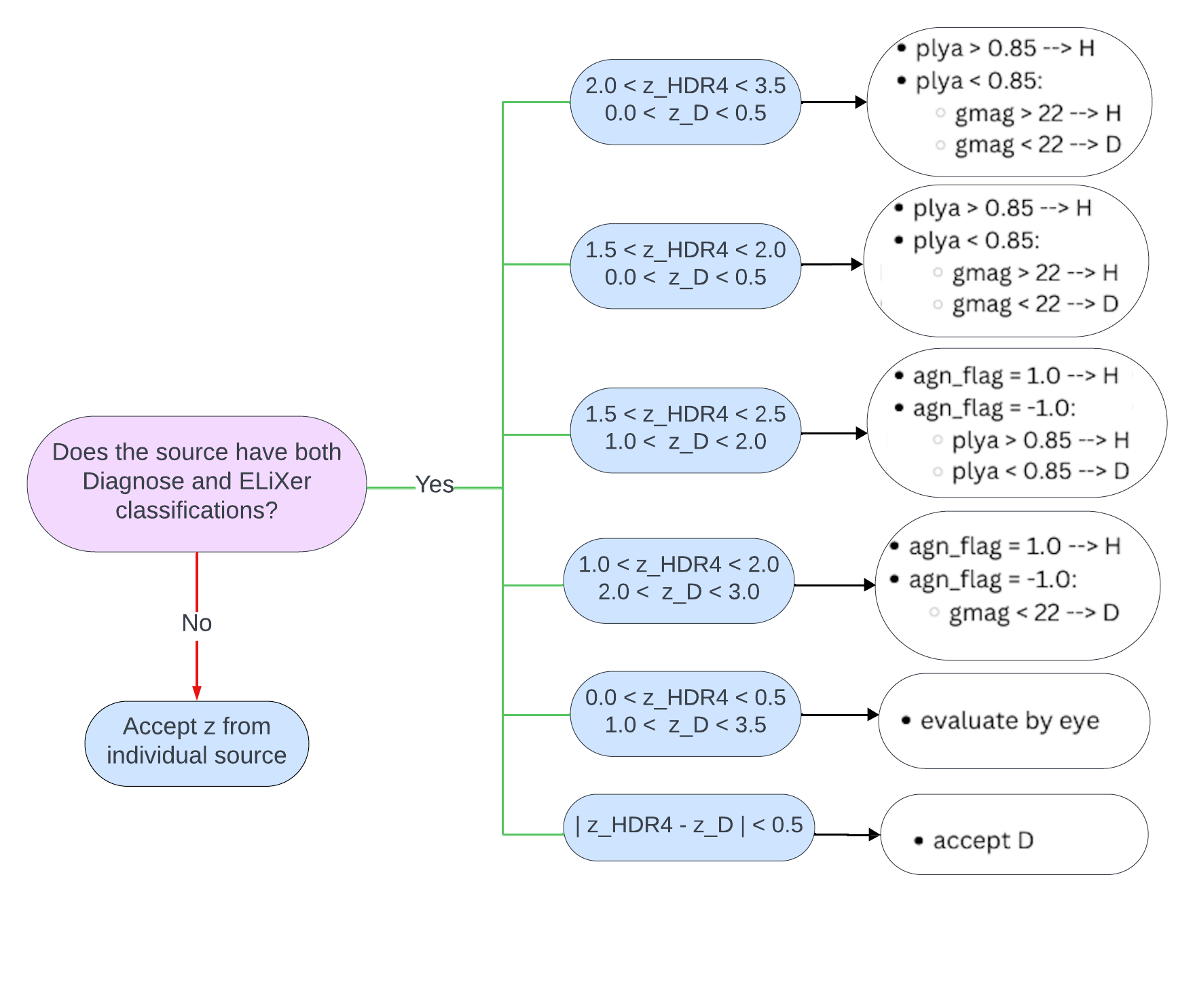}
  \caption{HETDEX-LOFAR Redshift and Classification Pipeline. The process begins with the LOFAR detection, followed by the extraction of HETDEX spectra. Once we have all spectra detected by HETDEX in the LoTSS field, we then run each spectrum through Diagnose and ELiXer to determine whether they have a classification and redshift from either source. If they have either a Diagnose or ELiXer redshift, we accept this value. If they have both, we follow a detailed examination of the difference in redshift between the two sources. The specifics of this examination are detailed in Appendix \ref{appendix:a}. If neither Diagnose nor ELiXer provide a redshift, we see if there is an archival value. If so, we accept this value. If not, the source remains unclassified. }
  \label{fig:hdr4_diagnose_disagree}
\end{figure*}

\begin{figure}[H]
  \centering
  \includegraphics[width=1\columnwidth]{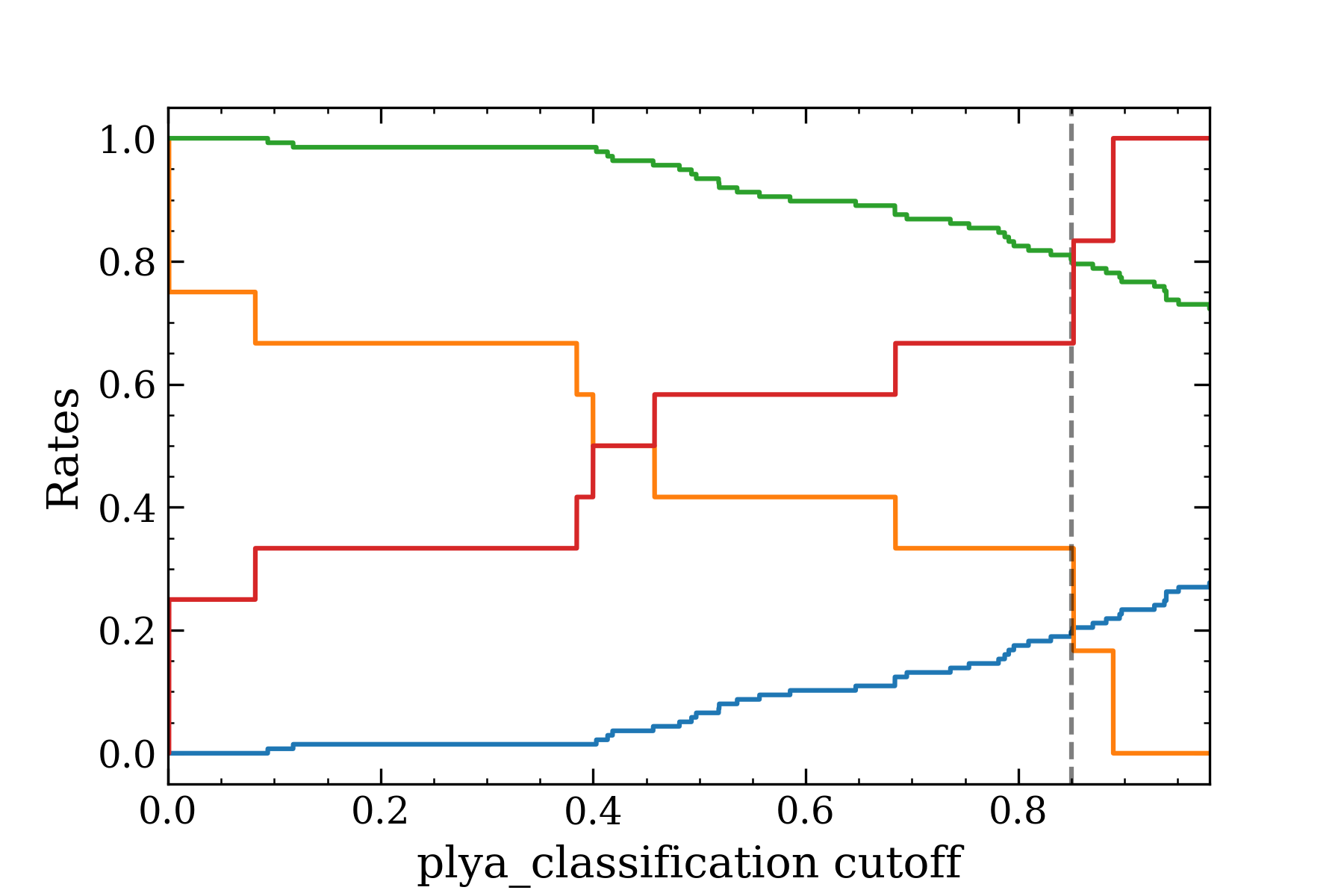}
  \caption{False positive (blue), false negative (orange), true positive (green), and true negative (red) rates as a function of `plya\_classifcation' cutoff. The grey dashed line represent a cutoff of 0.85.}
  \label{fig:plya_cutoff}
\end{figure}

There were 149 sources with $2.0 < z_{HDR4} < 3.5$ and $0.0 < z_{Diagnose} < 0.5$, which demonstrate the common issue of Diagnose identifying an emission line as \OII\ while HDR4 identifies the line as Ly$\alpha$. In order to distinguish which is the best fit redshift for this group, we utilized the HDR4 catalog output `plya\_classification' which represents the likelihood of the detected line is Ly$\alpha$. This criteria ranks each object from 0-1 with 1 being a high probability that the line is Ly$\alpha$. In order to utilize this probability, we determined a cutoff of `plya\_classification' = 0.85 using the HDR4 classification scheme that uses Diagnose redshifts for g-band magnitudes brighter than 22. Figure~\ref{fig:plya_cutoff} demonstrates how we determined this cutoff using false positive and false negative rates for different cutoffs.

To determine the rates, we examined each source's `plya\_classifcation' and g-band magnitude at incremental cutoffs between 0 and 1. If the `plya\_classifcation' was greater than the cutoff and the g-band magnitude was greater than 22, then we considered this a true positive identification. If the `plya\_classifcation' was greater than the cutoff and the g-band magnitude was less than 22, then we considered this a false positive identification. If the `plya\_classifcation' was less than the cutoff and the g-band magnitude was less than 22, then we considered this a true negative identification. If the `plya\_classifcation' was less than the cutoff and the g-band magnitude was greater than 22, then we considered this a false negative identification. The true positive rate was calculated as true positives divided by the sum of true positives and false negatives. The true negative rate was calculated as true negatives divided by the sum of true negatives and false positives. The false positive rate is 1-true positive rate, and the false negative rate is 1-true negative rate.

\begin{figure}[t]
  \centering
  \includegraphics[width=1\columnwidth]{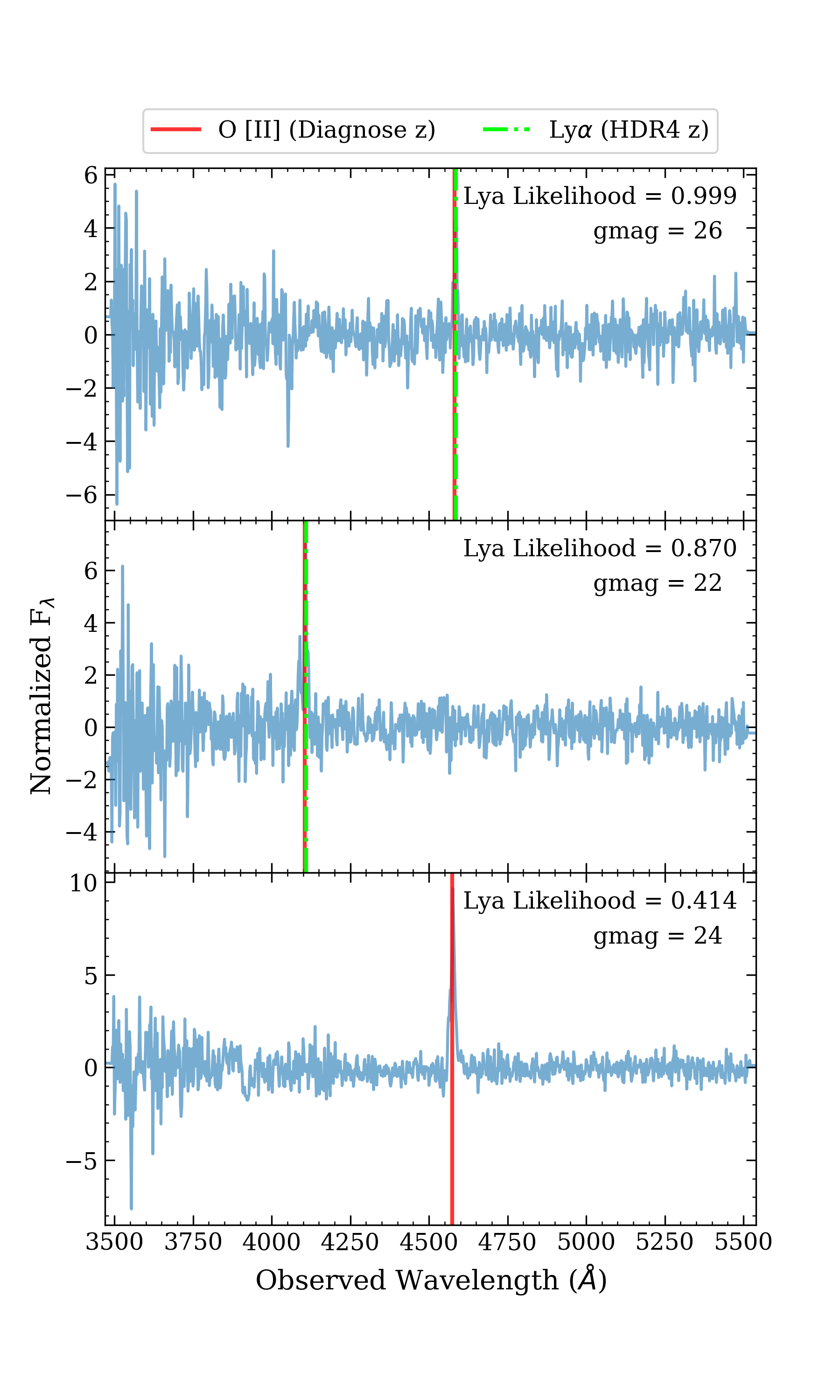}
  \caption{Three cases of redshift determination for $2.0 < z_{HDR4} < 3.5$ and $0.0 < z_{Diagnose} < 0.5$. The top plot shows the case where `plya\_classifcation' $> 0.85$ and the g-band magnitude is greater than 22, resulting in the use of the HDR4 redshift. The middle plot shows the case where `plya\_classifcation' $> 0.85$ but the g-band magnitude is bordering 22, resulting in the use of the Diagnose redshift. The bottom plot shows the case where `plya\_classifcation' $< 0.85$, resulting in the use of the Diagnose redshift.}
  \label{fig:lae_oii}
\end{figure}

After determining a cutoff for `plya\_classifcation', we then proceeded to examine each source in this redshift range by eye. Through this investigation, we determined three patterns to define which redshift to use. Figure~\ref{fig:lae_oii} shows the three example cases. Using these three patterns, we then determined that for sources that fall within $2.0 < z_{HDR4} < 3.5$ and $0.0 < z_{Diagnose} < 0.5$, if the `plya\_classifcation' $< 0.85$, we accept the Diagnose redshift. If `plya\_classifcation' $> 0.85$ but the g-band magnitude is less than 22, we accept the Diagnose redshift. If `plya\_classifcation' $> 0.85$ and the g-band magnitude is greater than 22, we accept the HDR4 redshift.

\begin{figure}[t]
  \centering
  \includegraphics[width=1\columnwidth]{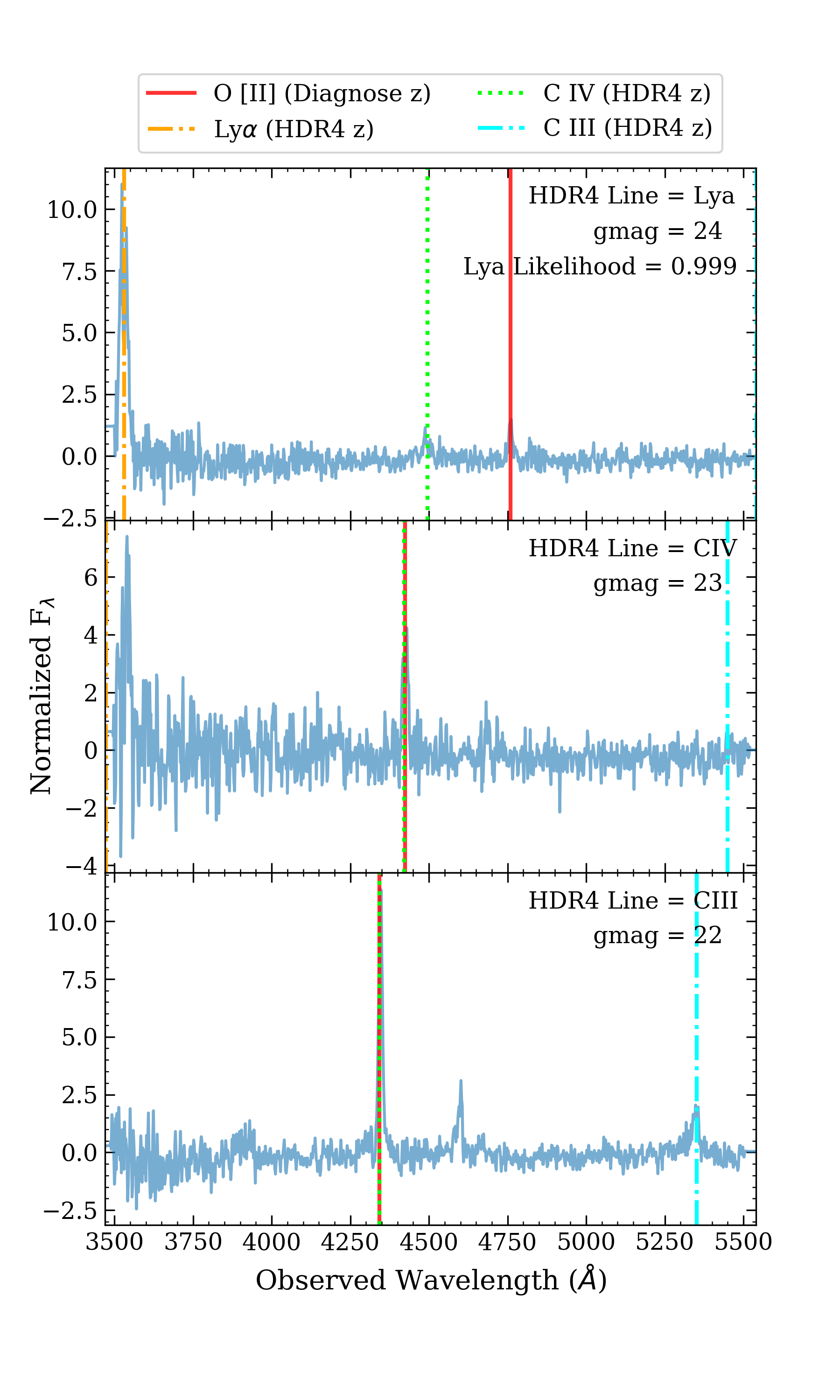}
  \caption{Three cases of redshift determination for $1.5 < z_{HDR4} < 2.0$ and $0.0 < z_{Diagnose} < 0.5$. The top plot shows the case where `plya\_classifcation' $> 0.85$ and the g-band magnitude is greater than 22, resulting in the use of the HDR4 redshift. The middle plot shows the case where the g-band magnitude is greater than 22, resulting in the use of the HDR4 redshift. The bottom plot shows the case where the g-band magnitude is bordering 22, resulting in the use of the Diagnose redshift.}
  \label{fig:other_oii}
\end{figure}

There were 20 sources with $1.5 < z_{HDR4} < 2.0$ and $0.0 < z_{Diagnose} < 0.5$. Some sources in this range also demonstrate the common issue of Diagnose identifying an emission line as \OII\ while HDR4 identifies the line as Ly$\alpha$. In order to distinguish which is the best fit redshift for this group, we utilized the HDR4 catalog output `plya\_classification' which represents the likelihood of the detected line is Ly$\alpha$ when HDR4 identifies the emission line as Ly$\alpha$. After investigating by eye, we determined three patterns to define which redshift to use. Figure~\ref{fig:other_oii} shows the three example cases. Using these three patterns, we then determined that for sources that fall within $1.5 < z_{HDR4} < 2.0$ and $0.0 < z_{Diagnose} < 0.5$, if the line identified is Ly$\alpha$ and `plya\_classifcation' $< 0.85$, we accept the Diagnose redshift. If the line identified is Ly$\alpha$ and `plya\_classifcation' $> 0.85$ but the g-band magnitude is less than 22, we accept the Diagnose redshift. If the line identified is Ly$\alpha$ and `plya\_classifcation' $> 0.85$ and the g-band magnitude is greater than 22, we accept the HDR4 redshift. If the line identified is not Ly$\alpha$, then we determine solely based on g-band magnitude. If the g-band magnitude is greater than 22, we accept the HDR4 redshift, and if the g-band magnitude is less than 22, we accept the Diagnose redshift.

\begin{figure}[t]
  \centering
  \includegraphics[width=1\columnwidth]{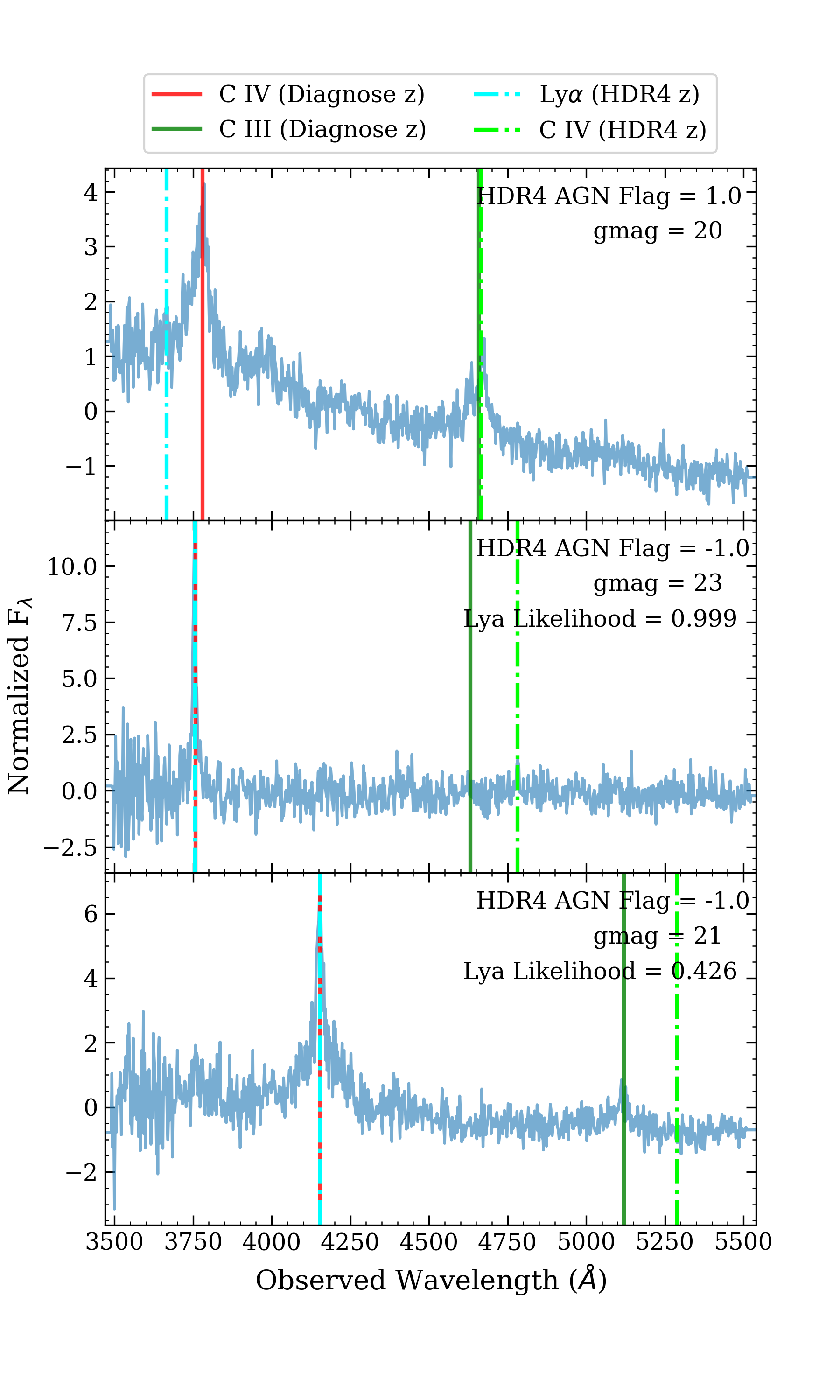}
  \caption{Three cases of redshift determination for $1.5 < z_{HDR4} < 2.5$ and $1.0 < z_{Diagnose} < 2.0$. The top plot shows the case where `agn\_flag' $= 1.0$, resulting in the use of the HDR4 redshift. The middle plot shows the case where `plya\_classifcation' $> 0.85$ and the g-band magnitude is greater than 22, resulting in the use of the HDR4 redshift. The bottom plot shows the case where `plya\_classifcation' $< 0.85$, resulting in the use of the Diagnose redshift.}
  \label{fig:agn_plya}
\end{figure}

There were 18 sources with $1.5 < z_{HDR4} < 2.5$ and $1.0 < z_{Diagnose} < 2.0$. In order to distinguish which is the best fit redshift for this group, we utilized the HDR4 catalog output `agn\_flag' which represents the confidence in the HDR4 AGN classification. A score of 1.0 is a confident AGN, 0.0 is a broadline source but unconfirmed AGN, and -1.0 means it is not an AGN. After investigating by eye, we determined three patterns to define which redshift to use. Figure~\ref{fig:agn_plya} shows the three example cases. Using these three patterns, we then determined that for sources that fall within $1.5 < z_{HDR4} < 2.5$ and $1.0 < z_{Diagnose} < 2.0$, if `agn\_flag' $= 1.0$, we accept the HDR4 redshift. If `agn\_flag' $= -1.0$ and `plya\_classifcation' $> 0.85$ but the g-band magnitude is less than 22, we accept the Diagnose redshift. If `agn\_flag' $= -1.0$ and `plya\_classifcation' $> 0.85$ and the g-band magnitude is greater than 22, we accept the HDR4 redshift.

\begin{figure}[t]
  \centering
  \includegraphics[width=1\columnwidth]{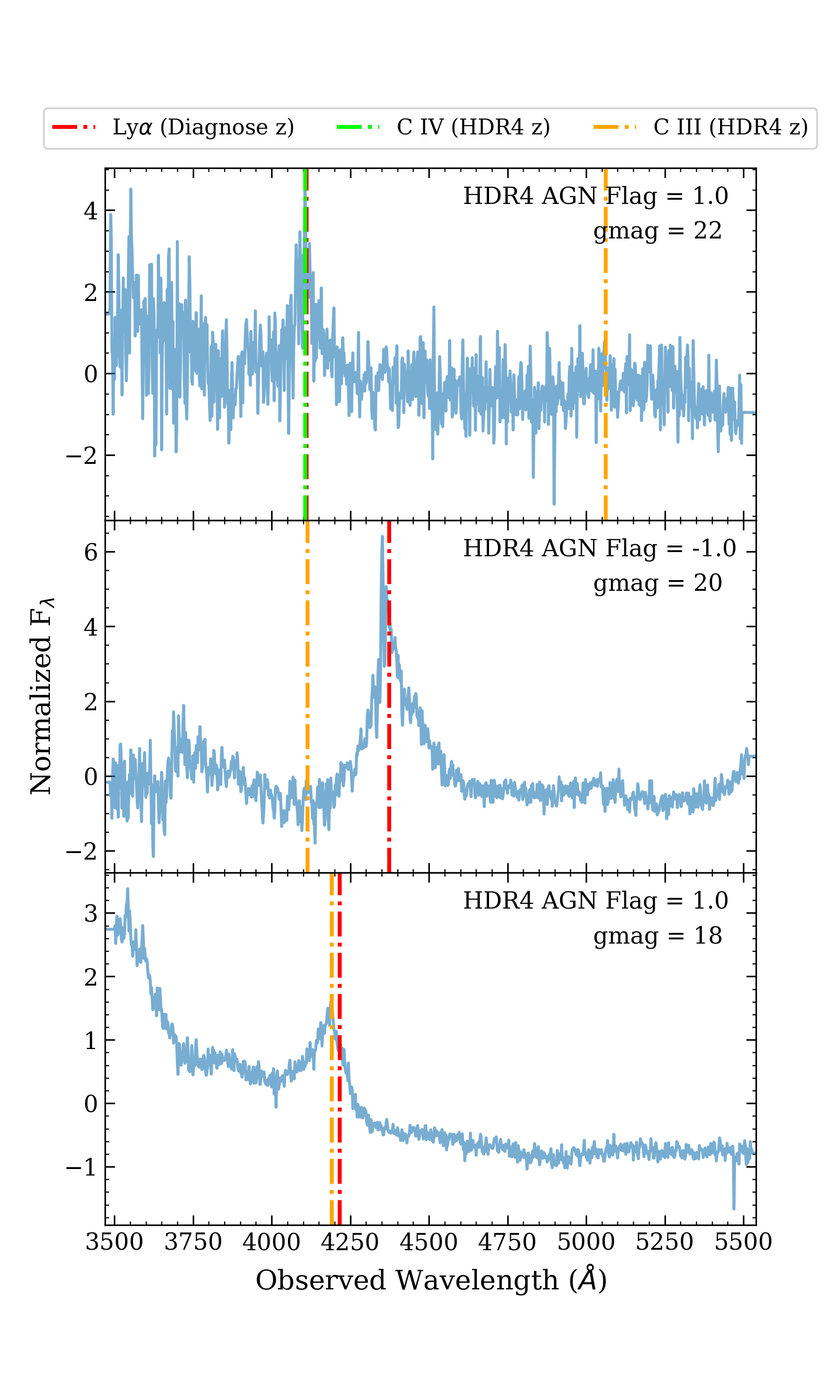}
  \caption{Two cases of redshift determination for $1.0 < z_{HDR4} < 2.0$ and $2.0 < z_{Diagnose} < 3.0$. The top and bottom plots show the case where `agn\_flag' $= 1.0$, resulting in the use of the HDR4 redshift. The middle plot shows the case where `agn\_flag' $= -1.0$ and the g-band magnitude is less than 22, resulting in the use of the Diagnose redshift.}
  \label{fig:agn_gmag}
\end{figure}

There were 10 sources with $1.0 < z_{HDR4} < 2.0$ and $2.0 < z_{Diagnose} < 3.0$. In order to distinguish which is the best fit redshift for this group, we utilized the HDR4 catalog output `agn\_flag' which represents the confidence in the HDR4 AGN classification. After investigating by eye, we determined two patterns to define which redshift to use. Figure~\ref{fig:agn_gmag} shows the two example cases. Using these two patterns, we then determined that for sources that fall within $1.0 < z_{HDR4} < 2.0$ and $2.0 < z_{Diagnose} < 3.0$, if `agn\_flag' $= 1.0$, we accept the HDR4 redshift. If `agn\_flag' $= -1.0$ and the g-band magnitude is less than 22, we accept the Diagnose redshift. If `agn\_flag' $= -1.0$ and the g-band magnitude is greater than 22, we accept the HDR4 redshift.

\begin{figure*}[t]
  \centering
  \includegraphics[scale=0.5,width=2\columnwidth]{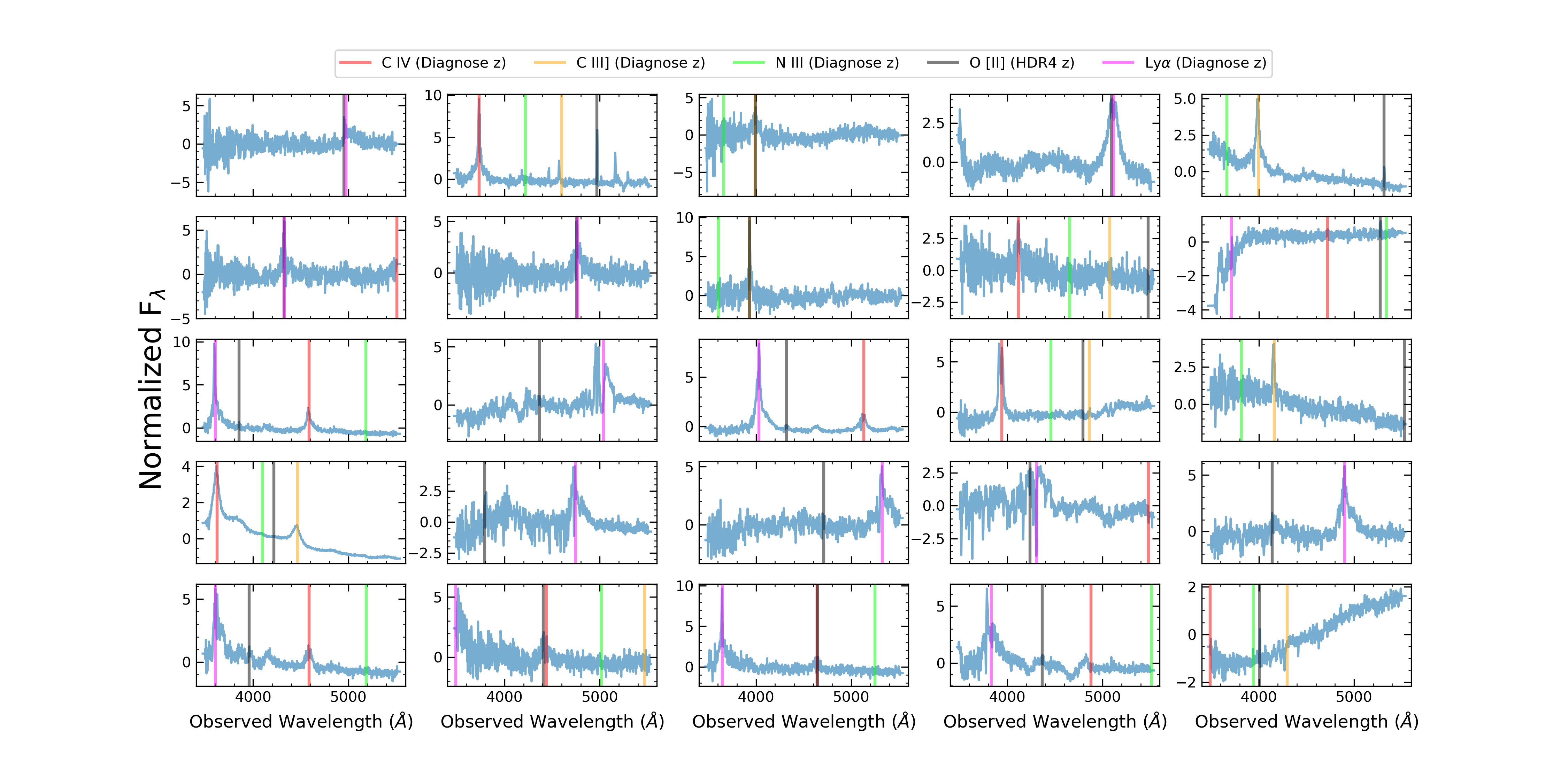}
  \caption{Plots of redshift determination for $0.0 < z_{HDR4} < 0.5$ and $1.0 < z_{Diagnose} < 3.5$. Each colored vertical line represents a different emission line location determined by either the Diagnose or HDR4 redshift. To classify each of these objects, we examined the line detected by HDR4, which was dominated by \OII\ and Ly$\alpha$. If by optical examination we did not find the HDR4 line at a significant feature, we then examined emission lines from Diagnose. In all of these cases, we found the g-band magnitude to be less than 22 and the Diagnose emission lines to be better representative of the spectra than the HDR4 determined lines.}
  \label{fig:hdr4_oii}
\end{figure*}

We investigated an additional category of 25 sources with $0.0 < z_{HDR4} < 0.5$ and $1.0 < z_{Diagnose} < 3.5$. These sources could not follow a specific set of criteria due to the large range of Diagnose redshifts. Figure~\ref{fig:hdr4_oii} shows the spectrum for each object in this group.

\end{document}